\documentclass[aps,prb,twocolumn,notitlepage,showpacs,superscriptaddress,10pt,footinbib,floatfix]{revtex4-2}% %DIF > 
\usepackage{graphicx}
\usepackage{amsmath}
\usepackage{bm}
\usepackage{amssymb}
\usepackage{color,soul}
\usepackage{ulem}  %ky for sout
\usepackage{amsfonts}%
\usepackage[caption=false]{subfig}
\usepackage{comment}
\usepackage{color,soul}
\usepackage[dvipsnames]{xcolor}
\usepackage[capitalize]{cleveref}
\usepackage{float}
\usepackage{tabularx} 
\usepackage{array}   
\newcolumntype{L}{>{$}l<{$}} 
\newcolumntype{R}{>{$}r<{$}} 
\newcolumntype{C}{>{$}c<{$}} 
\usepackage{comment}
\usepackage[capitalize]{cleveref}
\newcommand\thao[1]{\textcolor{red}{(TN: #1)}}

\newcommand\blue[1]{{\color{blue}#1}}

\newcommand\cacro{CaCrO$_3$}

\newcommand\3{${_3}$}

\begin{document}
%\begin{frontmatter}
%\title{Electric field control the magnetism in bilayer VI_{3}}%
\title{
\textit{Ab initio} prediction of  anomalous Hall effect \\
in antiferromagnetic CaCrO$_3$
} %Force line breaks with \\



\author{Thi Phuong Thao Nguyen}
\author{Kunihiko Yamauchi}
\affiliation{%
 Institute of Scientific and Industrial Research, Osaka University, 8-1 Mihogaoka, Ibaraki, Osaka, 567-0047, Japan
}%
\affiliation{% 
Department of Precision Engineering, Graduate School of Engineering, Osaka University, 2-1 Yamadaoka, Suita, Osaka 565-0871, Japan
}%

\date{\today}% It is always \today, today,

\begin{abstract}
%\ky{Correct the tense overall in the manuscript!! Use one of three: (1) It has been found/discussed by someone else in the past [for previous works]. (2) This calculation was done by us [past for what we did]. (3) It is found that AHC is induced by SOC [present fact].}

%\sout{Anomalous Hall conductivity was evaluated in collinear antiferromagnetic \cacro\ by using density-functional-theory calculations. We found that the anomalous Hall conductivity is finite in the C-type collinear antiferromagnetic phase \cacro\, even though the net magnetization is almost zero. The key ingredient to anomalous Hall components comes from magnetic order parameters belonging to the same irreducible representation in the non-symmorphic space group. We identify that the origin of nonzero Berry curvature is the spin-splitting bands at the gapped nodal lines by spin-orbit coupling near the Fermi energy. }
%
%\blue{

While the anomalous Hall effect takes place typically in ferromagnets with finite magnetization, large anomalous Hall conductivity in {\it noncollinear} antiferromagnetic systems has been recently observed and attracted much attention. 
In this study, we predict the anomalous Hall effect in perovskite CaCrO$_3$ as a representative of {\it `collinear'} antiferromagnetic materials. Our result shows that the C-type antiferromagnetic ordering generates the sizable anomalous Hall conductivity. % whose value is comparable to that in ferromagnetic ordering. 
Based on symmetry analyses, we show that 
%it is essential that 
the antiferromagnetic order parameter belongs to the same irreducible representation as the ferromagnetic order parameter in  the nonsymmorphic space group, allowing the non-vanishing Berry curvatures in $k$  space.  
%(Caused by non-vanishing Berry curvature in antiferromagnetic material with certain magnetic space group symmetry, anomalous Hall conductivity can be enhanced by the spin-splitting phenomena in the band structure instead of the net magnetization.)
By performing first-principles density-functional theory calculations, 
we find that the Berry-curvature {\it `hot spots'} lie along the gapped nodal lines where spin-orbit coupling induces the spin splitting of Cr-3$d$ bands near the Fermi energy and enhances the anomalous Hall effect in CaCrO$_3$. 
%}

%In contrast to the conventional anomalous Hall effect in ferromagnetic or non-collinear antiferromagnetic states the collinear AFM coupling with ferromagnetic components in nonsymmorphic orthorhombic crystal structure reveals the manifestation of Hall conductivity. %\ky{not clear at all}\ky{Rewrite abstract in the end of correction}
  
\end{abstract}

\maketitle

\section{Introduction}
%\brown{\textbf{Anomalous Hall conductivity }
%\begin{itemize}
%    \item Large AHE in Mn$_3$Sn and Mn$_3$Ge was theoretically found by K\"{u}bler~\cite{Kubler2014, Kubler2017}. Experimentally observed by Nakatsuji group, analyzed by Suzuki. 
%    \item Spin chirality discussion (Tokura group) VS orbital-driven Berry phase discussion (Kontani group). 
 %   \item Unconventional Anomalous Hall Effect in the Canted Antiferromagnetic Half-Heusler Compound DyPtBi 
%\end{itemize}}

The anomalous Hall effect (AHE) is traditionally considered to be proportional to net magnetization and therefore appears only in ferromagnetic (FM) materials~\cite{luttinger1954}.
Nevertheless, recent studies have revealed that large anomalous Hall conductivity (AHC) emerges in Kagome antiferromagnetic (AFM) materials such as Mn\3Ge and  Mn\3Sn regardless of their small net magnetization~\cite{ Chen2014, Kubler2014, Yang_2017, Zhang2017, Nakatsuji2015_mn3sn, Kiyohara2016_mn3ge, Nayak2016, Yoon2017, Chen_Nakatsuji2021}.
%The anomalous Hall effect typically appears in ferromagnets with finite magnetization. \sout{ and has been rarely seen in antiferromagnetic materials.} Recently, large anomalous Hall conductivity (AHC) in noncollinear antiferromagnetic (AFM) has been observed.
%\ky{Avoid duplicate text and summarize it shortly} \sout{Anomalous Hall conductivity has been predicted in noncolinear antiferromagnets} 
%After the first prediction, AHC was experimentally recognized in MnGe\3 and  MnSn\3 metals~\cite{}. 
%In these works, 
Such a large AHC in antiferromagnets  may work as a readout for a spintronic device accompanied by their ultra-fast spin dynamics and insensitivity to external magnetic fields, whereas only a few AFM materials have been studied in this context~\cite{Shindou2001, Shi2018_AHE_Ti2MnA, Smejkal2020_ruo2, Samanta2020_SrRuO3, Li2019_QAHE_AFM}.% up to now. 
Suzuki {\it et al.} have investigated the magnetic symmetry by introducing a magnetic cluster extension; %to discuss the anomalous Hall effect
the noncollinear AFM structure in Mn$_3$Sn is characterized by the cluster octupole moment, which belongs to the same irreducible representation as the  collinear ferromagnetic ordering. This explains why the noncollinear antiferromagnetic order can induce an AHC despite its vanishingly small dipole magnetization~\cite{Suzuki2017}.
The same approach has been later applied to antiperovskite manganese nitrides such as Mn$_3$PtN, where %different types of noncollinear antiferromagnetic ordering can emerge, leading to the nonzero AHC~\cite{Boldrin2019, Zhou2019, HuyenKY2019, Gurung2019}. %The study of multipole expansion for magnetic structures
%In either case, noncollinear magnetism is necessary to obtain the AHC in AFM materials. 
%\ky{Not clear. Explain noncoplanar and coplanar spin configurations to distinguish them.}
%In Mn3PtN, similar magnetic symmetry analysis has been done. T
their magnetic octupole noncollinear AFM states allow for inducing the large AHC as well~\cite{HuyenKY2019}.  %, exhibit the AH conductivities comparable to those in ferromagnetic states of Fe and Co in size. 

%\ky{here we discuss hot spot to enhance AHC. need to shorten following text taken from vanderbilt's book}
Recent theoretical development  provides the topological formulation of the intrinsic AHE in terms of the Berry phase  associated with the Bloch wave functions in solids~\cite{vanderbilt_2018}.
%\sout{In Mn\3Sn, the orbital-derived Berry phase with the coplanar spin structure has been considered to cause the AHE~\cite{Tomizawa_Kontani2009}.} 
While the AHC can be calculated by integrating the Berry curvature in the Brillouin zone, the computation is practically  demanding 
because the Berry curvature %is necessary to integrate over the Brillouin zone and it is 
in many cases a very sharply varying function of $k$ vector. It  often strongly depends on contributions from a few {\it “hot spots”} in the $k$-space where the spin-orbit-coupling (SOC) causes anti-crossing between bands near the Fermi energy. 
Such features often require an extremely dense $k$-point sampling, and for this reason, the Wannier functions are employed to interpolate the band structure and the wave functions~\cite{vanderbilt_2018}.  %\ky{thao, cite vanderbilt book}
%\sout{One of the famous example of AHC in AFM is Mn$X$\3($X$=Ga, Ge, Sn, Ir).} 
%
In Mn\3Sn, the Berry curvature stemming from the Weyl points enhances AHC in the absence of net magnetization~\cite{Chen_Nakatsuji2021}.%~\cite{Kubler2014, Kubler2017}.%\ky{thao, cite a paper}. 
In contrast, in Mn\3PtN,  %exhibits the AH conductivities comparable to those in size in the ferromagnetic states of Fe and Co. It has been shown that 
the Berry curvatures spread around the Fermi surfaces in the broad Brillouin-zone region, coming from the band splitting due to the SOC, dominantly contributes to the AHC~\cite{HuyenKY2019}. %, while the locally divergent Berry curvature produces only a small contribution to the AH conductivity after considering the band summation and BZ integral~. % It opens a viewpoint for a relation between topol- ogy and macroscopic phenomena in noncollinear AFM~\cite{HuyenKY2019}.

Very recently, Naka {\it et al.} have theoretically examined the possibility of AHE in perovskite transition-metal oxides~\cite{Naka2022_AHC}. 
%that certain collinear antiferromagnetic materials can exhibit the AHE 
%by a tight-binding-model calculation %\ky{the citation looks wrong. 21 is for spin current, isn't it? 22 has no information of paper. Do you mean arxiv paper? }\thao{21 for spin current, 22 for arxiv paper, I correct ref. Naka2022}
It has been predicted that the coexistence of the GdFeO\3-type structural distortion and the collinear AFM configuration gives rise to the AHE %even though the net magnetization is zero.
by means of the Hubbard model, whereas such unusual AHE has not been experimentally studied so far.   
One of the candidate materials proposed %in their study 
is 
%\sout{The conduction behavior of transition metal oxides is essentially related to the magnetic ground state: the FM order is associated with metallic conductivity, and the AFM order is with insulating conductivity.}\ky{Is it always so?? you had better explain CMR effect in half-doped manganites.} 
\cacro, which is a rare example of a metallic and antiferromagnetic transition-metal oxide.   % with a GdFeO\3-type structural distortion. %\ky{what kind of lattice distortion?} 
\cacro\ crystallizes in orthorhombic $Pbnm$  %\ky{you mean Pnma? Don't make such a mistake} 
(alternative setting of $Pnma$) perovskite structure. 
Although CaCrO$_3$ was previously reported to show semiconducting~\cite{GOODENOUGH1968_cacro3} or insulating~\cite{Zhou2006_cacro3} properties with the AFM order, recent works have reported its metallic conductivity below the N\'{e}el temperature, $T_{\rm N} = 90$K~\cite{E_Castillo-Martinez_J_Solid_State_Chem_181_895_2008, Weiher1971_cacro3}.
Powder neutron-diffraction analysis and $\mu$SR measurement have revealed that the AFM spin structure is C-type AFM (C-AFM) in which Cr spins order antiferromagnetically in the $ab$ plane but ferromagnetically along the $c$ axis~\cite{Komarek2008_cacro3, Ofer2010_CyAFM_muSRmeasure}. 
%\ky{that is important. explain better in another paragraph.} \sout{Owning \red{to} a similar orthorhombic structure with AFM nature but with the metallic ground state, \cacro\ is expected to show anomalous behavior in its crystal structure and metallic conductivity.}\ky{you should write motivation of this study here.}
Owing to the orthorhombic lattice distortion with the metallic collinear AFM ground state, \cacro\ is a suitable playground to realize the AHE with no need to consider the complex magnetism such as spin chirality and/or magnetic multipole configurations. In this study, we perform DFT calculations to evaluate the AHC in \cacro\ 
and discuss  % the singularity spots of the Berry curvatures in \cacro\ and analysis of the magnetic symmetry   
%to understand 
its microscopic mechanism %of the AHE 
along with the magnetic symmetry analysis.

%%%%%%%%%%%% -----Motivation of this work---%%%%%%%%%%%%%
%\sout{Here we are taking a step further by considering a collinear antiferromagnetic oxide as a playground for the anomalous Hall effect.}\ky{we are doing very new work, right?} 
%DFT simulations reveal the unexpected presence of AHC in the C-type AFM \cacro\ with vanishing net magnetization.\ky{this is conclusion. don't write it here.} We found that the calculated AHC in the C-type AFM phase is finite and comparable with that in the ferromagnetic phase. Using the Wannier-function interpolation technique, we evaluated the Berry curvatures and discussed the microscopic origin of the large AHC. \ky{don't write it here.} 

%\thao{Don't need this part in the short paper}The paper is organized as follows; we first focus on the electronic and magnetic stability of \cacro\ in ~\cref{ssec:properties}. Then we show the role of the nonsymmorphic space group in the AHC via symmetry analysis in ~\cref{ssec:symmetry}. In~\cref{ssec:ahc}, we discuss the existence of AHC. Then, we address the origin by examining the hot spot which primarily contributes to AHC in the Brillouin zone via the Berry curvature and the Fermi surface in ~\cref{ssec:mechanism}.  Finally, we draw conclusions in ~\cref{sec:conclude}.
%\thao{2022/06/29 (4 hours)}

\section{\label{sec:method}Methodology} 
Density-functional theory (DFT) calculations were performed by using VASP \cite{vasp} and QUANTUM ESPRESSO \cite{quantum_espresso} packages. The generalized gradient approximation with the parametrization of Perdew, Burke, and Ernzerhof (GGA-PBE) \cite{gga} was used for the exchange-correlation functional. %\ky{not function but functional!}. 
The calculations were carried out by following steps. First, the atomic structure was optimized with C-AFM configuration until the atomic force become lower than 0.001eV/{\AA} using VASP code with the PAW  pseudopotential~\cite{paw} and then the electronic structure and the magnetic stability were investigated.  %[The pseudo-potentials in the projector augmented-wave method~\cite{paw} was used. 
Secondly, another self-consistency calculation was performed by using QUANTUM ESPRESSO code %After checking the convergence of the calculation, 
with the fully relativistic ultrasoft pseudopotentials~\cite{PSLIBRARY}. The kinetic energy cutoff of 80 Ry and 800 Ry were used for the plane-wave basis set and charge density, respectively. 
%\sout{The self-consistency calculation was performed using a k-point grid of 8$\times$8$\times$6  with a Methfessel-Paxton smearing width of 0.02 Ry.} 
A 12$\times$12$\times$10 {\it k}-point mesh was taken in the Brillouin zone  with the tetrahedron method for integration. 
%\ky{separate computational process for DFT and Wannier. Don't mix two. }
%\ky{Thao, you didn’t write in what process you used VASP and QE. You must provide enough information to readers to reproduce your results. Write down what you really did here. I also remenber that you used ultrasoft psuedo potential. Explain it.}\thao{I added more information above}
Thirdly, we constructed maximally-localized Wannier functions by using WANNIER90 code~\cite{wannier90} that is interfaced with the QUANTUM ESPRESSO code. 
%The AHC was evaluated  with the uniform k-point mesh of 120$\times$120$\times$100 with the adaptive k-mesh refinement of 5$\times$5$\times$5 for the absolute values of Berry curvature larger than 10 {\AA}$^2$.
%The 3$d$ orbitals for Cr atoms are included for the Wannier interpolation scheme. 
The Wannier functions were built by projections of Bloch wavefunctions onto the localized Cr-3$d$ orbital basis;  there are 40 $d$ orbital states in the C-AFM unit cell (4 f.u.). These atomic orbital functions must be carefully chosen otherwise the maximally localization process of the Wannier functions does not converge well. 
We set up the local ($x, y, z$) axes in the CrO$_6$ octahedron by considering the Cr-O bond directions to satisfy the condition:  $z$//long bond, $x$//short bond, and $y$//middle bond as possible. The angular functions, $3z^2-r^2$, $zx$, $yz$, $x^2-y^2$, and $xy$, were defined in the local coordinates for the Wannier projection. 
After %obtaining the self-consistent potential \ky{what is it?} with 1440 \textbf{k} points
 the Wannier functions were maximally localized, we finally calculated the Berry curvature and the AHC with 120$\times$120$\times$100 $k$-point mesh 
%several larger sets of k points in order to achieve the convergence for AHC values.  When Berry curvature is larger than 10 {\AA}$^2$ at a certain k point, we construct a finer mesh by adding adaptive k-mesh refinement around it. AHC value reaches convergence under the evaluation with a uniform k-point mesh of 
  with 5$\times$5$\times$5 adaptive mesh refinement for the turbulent regions.

%\thao{Must check the theory of Berry Curvature on the Fermi Surface by Haldane~\cite{Haldane2004}.}

\section{\label{sec:results}Results and discussions} 

\subsection{\label{ssec:properties}Electronic and magnetic properties} 

\begin{figure}
    \centering
    \includegraphics[width=0.9\columnwidth]{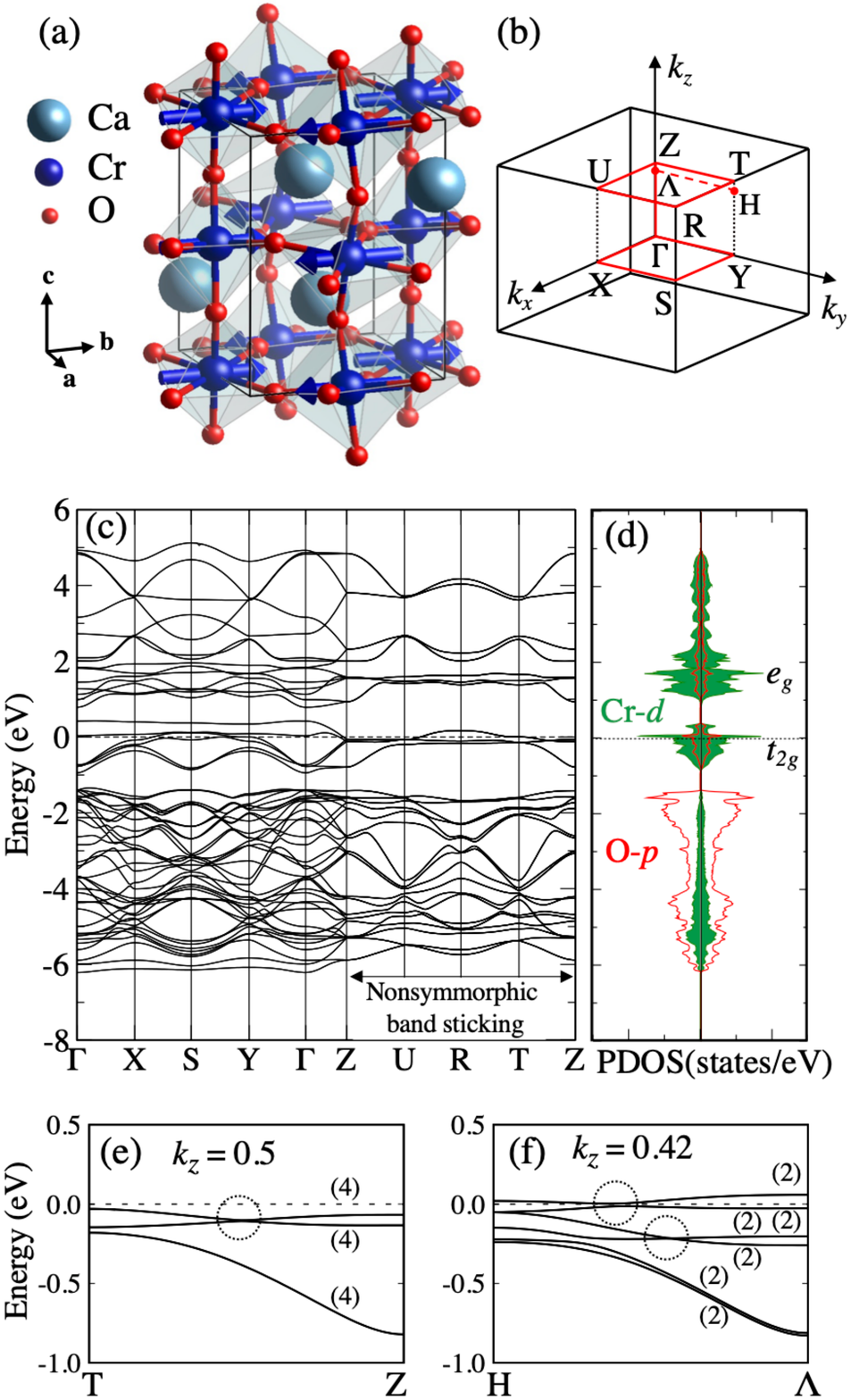}
    \caption{%\ky{Add pannels (e) and (f) ASAP!} 
(a) Orthorhombic $Pbnm$ crystal structure of \cacro\ with C-AFM spin configuration with Cr spins set parallel to the $y$ axis. (b) First Brillouin zone with the high-symmetry points. (c) Calculated band structure  with C-AFM configuration without SOC. (d) Corresponding DOS projected on Cr$-d$ (filled green) and O$-p$ (red) orbital states. (e)(f) %$t_{2g}$
    The band structure near the Fermi energy along the T-Z line and the ${\rm H}$-$\Lambda$ lines at $k_z$=0.42, respectively. The number in parentheses is the degeneracy for the band. Fermi energy is set at zero.%\ky{Don't create names such as W and L. Use $\Lambda1$ for (0 0 0.4) and $H_1$ for (0.5 0 0.4). Besides, the direction in (f) is opposite to definition in (b)! }
    %\ky{$y$ in $k_y$ is missing in (b).}
    %\thao{Update BZ, please check}
    }
    \label{fig:structure} 
\end{figure}

%\cacro\ shows an orthorhombic distorted GdFeO\3 structure. 
Figure \ref{fig:structure} (a) shows the $Pbnm$ crystal structure of \cacro\ with the orthorhombic Brillouin zone %used in the calculation %are shown in ~\cref{fig:structure}(a)-(b). 
%The lattice constant are $a = 5.33343${\AA}  $b = 5.47226${\AA}  $c=7.54876${\AA} (Material-Project id: mp-20772).\ky{write optimized lattice parameters within four decimal digits.} 
%\sout{The optimized lattice constants are %$a = 5.2873${\AA},  $b =  5.3567${\AA}  $c=7.4985${\AA}. listed in Tbl.II.} 
and the calculated electronic structure. %is shown in ~\cref{fig:structure}(c)-(d). 
The metallic state is clearly
exhibited by the 2/3-filled Cr-$t_{2g}$ state crossing the Fermi energy. Since the $t_{2g}$ level is located away from O-$p$ level and the three-fold degeneracy is not completely lifted due to the lack of strong Jahn-Teller distortion,  the $t_{2g}$ state has rather localized character and form the flat bands in the vicinity of the Fermi level. It also appears as a sharp peak, the so-called van Hove singularity, in the DOS, being responsible for the high electric conductivity.   % right above the Fermi level.  
In contrast, the $e_g$ orbital state shows a delocalized character as strongly hybridizing with O-$p$ state. The $e_g$-$p$ bonding state lies in a wide energy range below $t_{2g}$ state and the anti-bonding state spreads above $t_{2g}$ state. The trend is consistent with the previous DFT works.\cite{Streltsov_cacro3_dft, Liu2011_cacro3_dft}  

The nonmagnetic $Pbnm$ space group has eight symmetry operations: 
%\ky{make them italic + I changed the notation; follow this} 
\{$E$, $I$, $C_{2x}+(\frac{1}{2} \frac{1}{2} 0)$, $C_{2y}+(\frac{1}{2} \frac{1}{2} \frac{1}{2})$, $C_{2z}+(0 0 \frac{1}{2})$, $m_x +(\frac{1}{2} \frac{1}{2} 0)$, $m_y+(\frac{1}{2} \frac{1}{2} \frac{1}{2})$, $m_z+(0 0 \frac{1}{2})$\}, where $E$, $I$, $C$, and $m$ denote  identity, inversion, rotation, and mirror operations, respectively.  
%\ky{replace $\sigma$ by $m$ entirely in the paper. $\sigma$ must be used only for AHC.} \thao{done]
Three screw and two glide symmetries involving fractional translations manifest the nonsymmorphic space group. 
%Nonsymmorphic symmetries combine a fractional lattice translation with either a mirror reflection (glide plane) or a rotation (screw axis), resulting in a band-folding with crossings at the Brillouin zone (BZ) boundaries that are protected against hybridization.(npj Quantum Mater. 7, 31 (2022)) 

\begin{figure}
    \centering
    \includegraphics[width=0.9\columnwidth]{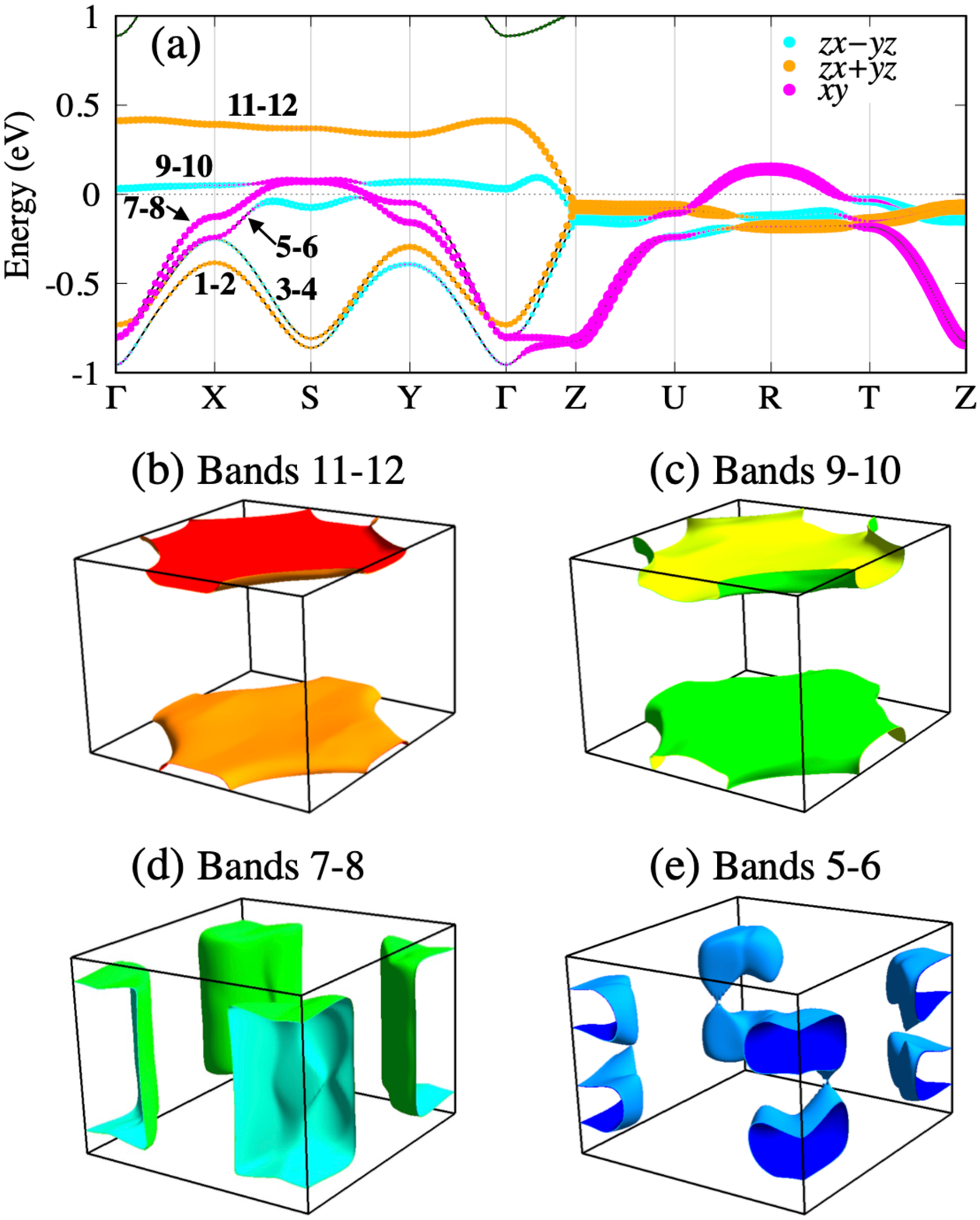}
    \caption{(a) %\ky{label 1-2, 3-4 bands, too} 
     Wannier-interpolated band structure under the C-AFM configuration projected onto Cr-$d$ orbital states without taking into account of SOC. The band number reflects the AFM spin-degeneracy. % under to the AFM order. 
    %Projected band structure on $t_{2g}$ orbitals in $C_y$ AFM configuration along the standard high-symmetry path. 
    The colors highlight %the bands represent 
    the following $d$-orbital components; magenta denotes $d_{xy}$, orange denotes $d_{zx+yz}$, and cyan denotes $d_{zx-yz}$ states. The radii of circles are proportional to the weights of corresponding orbital states. Fermi energy is set at zero. %(d) projected density of state (PDOS) on Cr\textit{-d} and O\textit{-p} orbital states.  
  (b-e) Calculated Fermi surfaces with the corresponding band numbers in the orthorhombic Brillouin zone.  
%  \ky{What do you mean by projected band structure? Is it Wannier result? If so, write it.}\thao{It's not Wannier projected bands. I modified the caption but I need a discussion at this point.}
}
    \label{fig:fermisurface} 
\end{figure}

The nonsymmorphic group operations lead to band degeneracies, so-called {\it ``band sticking''}, across an entire face of the Brillouin zone~\cite{Quan2022_bandsticking, Leonhardt2021_bandsticking}.  %and FSs pierce this wall of degeneracies.
%These nonsymmorphic symmetry operations induce band sticking across an entire surface of the BZ.
Figure \ref{fig:structure} (c) illustrates the band sticking effect in the  $k_z=0.5$ plane. 
Under the C-AFM configuration, 
all the bands are spin-degenerate at the high-symmetric $k$ points and the additional band-sticking  degeneracy is imposed at the Brillouin-zone surfaces. 
Therefore, two-fold 12 $t_{2g}$ bands become four-fold 6 bands in the Z-U-R-T plane (also see ~\cref{fig:fermisurface} (a)). 
%AFM spin configuration, together with the band-sticking effect, enforces the four-fold degeneracy bands on the $k_z=0.5$ plane.
%The description of band degeneracies in orthorhombic crystals 
%The degeneracy effect in the orthorhombic system has been summarized in Ref.~\onlinecite{Leonhardt2021_bandsticking}. 
%\blue{
The sticking effect causes a narrow bunch of bands just below the Fermi energy along the T-Z line at the $k_z$=0.5 plane (\cref{fig:structure} (e)). Along the H-$\Lambda$ line at the $k_z$=0.42 plane, they are split into two groups and the upper-lying bands manifest   the band crossing exactly at the Fermi energy (\cref{fig:structure} (f)), forming the nodal lines that will be important later in a following section.  
%}
%This work focuses on the band characteristic as well as the effect of band splitting by SOC. 
%The spin-splitting effect induced by SOC will be discussed in next section. 
%
%The electronic structures calculation suggests that the $t_{2g}$ orbitals of Cr significantly contribute near Fermi surfaces (FSs). 
%As shown in 
Figure \ref{fig:fermisurface}   shows the  band structure projected onto Cr-$d$ orbital states and the Fermi surfaces  obtained by using the Wannier-function interpolation. %There are twelve bands near the Fermi level and % 
The $t_{2g}$ orbital state are split into $xy$ and $(zx\pm yz)$ orbital states; 
the latter is a linear combination of two orbital states at neighboring Cr sites. 
%character mainly of $d_{zx-yz}$, $d_{zx+yz}$, and $d_{xy}$ components. Here we note that the local $z$ axis is set as an out-of-plane direction. 
Among them, 5-6th and 7-8th bands form the hole Fermi-surface pockets around the RS line and 9-10th and 11-12th bands form the lotus-flower shaped flat electron Fermi surfaces near the $k_{z}=0.5$ plane.  

The magnetic stability was examined by comparing the total energy between several magnetic configurations. %. The result is as follows
It is found that C-AFM order shows the total energy lower than those of other spin configurations: %\sout{ferromagnetic (FM), G-type AFM, and A-type AFM by 3.1, 7.7, and 1.2 meV/f.u.}\red{
ferromagnetic (FM), A-type AFM, and G-type AFM by 209.2, 59.4 and 103.9 meV/f.u, respectively.
%, being consistent with the experimental observation. %study that \cacro\ is an antiferromagnetic metallic transition-metal oxide with a C-type magnetic structure. Besides, 
Taking into account SOC, it is also found that C-AFM order shows the magnetic easy axis along the global $y$ direction. The energy difference is 0.76 meV/f.u. and 0.98 meV/f.u. with respect to the energy with the $x$ and the $z$ spin directions, respectively. %The result is consistent with the experimental observation.\cite{Ofer2010_CyAFM_muSRmeasure, Komarek2008_cacro3}  
%ground state reported as C-AFM configuration with local spins ordered along the $y$ direction.
%\ky{any experimental study that supports Cy AFM order? if so, cite it here.}
%with S$||$010 (C$_y$) is slightly more stable than one with S$||$100 (C$_x$) and S$||$001 (C$_z$) by xxx and xxx meV/f.u, indicating that the easy spin aligns in the \textit{y} axis.
%\ky{here you need to explain why you calculate AHC for Cx and Cy-AFM.}
Since the magnetic anisotropy energy is very small, hereinafter we will examine the AHC in C-AFM configuration with spins ordered along the $x, y,$ and $z$ axes to compare the values.  Then we will focus on the C-AFM order with the $y$ spin orientation to investigate the origin of the AHC in more detail.

\subsection{\label{ssec:symmetry}Magnetic symmetry analysis}
%\thao{Add KY table here. Ask KY to write the detail of this part.}
%\ky{Naka san defined it in his arxiv paper: "For example, when a C-type AFM pattern for the x axis spin moment is realized, owing to the symme- try, projections to the other axes must show ferromagnetic (F ) moments along the y axis and an A-type AFM pattern along the z axis; it is written as CxFyAz [30, 42, 43]. For simpli- fication, we will represent each pattern by the major compo- nent with the largest projected spin moments, e.g., as Cx-type AFM state (see below). ". Please follow him.}
%\thao{Komarek's ACrO3 magnetic space group discussion. Also check Solovyev1996 and Naka2022 \cite{Solovyev1996_lacro3, Komarek2008_cacro3, Naka2022_AHC}: Generally, there exist four different types of magnetic ordering schemes within representation theory for k = 0 and space group Pbnm (considering only the Cr4+ sites): The GxAyFz, AxGyCz, CxFy Az, and FxCyGz types of magnetic ordering are associated with the four different irreducible representations. Because the predominating magnetic mode in CaCrO3 is of the Cy type, the magnetic structure of CaCrO3 is of the Fx Cy Gz type according to the irreducible representation. However, even in a measurement at the high-flux diffractometer D20 yielding high statistics, it was not possible to detect any other intensities which would indicate additional magnetic components. The Fx component, however, is in perfect accordance with the observation of weak ferromagnetism in the susceptibility}

%\thao{Introduce Cy pattern here before magnetic discussion}
When the SOC is involved under the magnetic configuration, 
the crystalline symmetry is lowered   and 
we must %consider 
  %align in a specific direction
introduce the magnetic space group considering the specified spin direction. 
In \cref{table:symmetry}, we show the transformation rules in nonmagnetic space group $Pbnm1$' considering FM (F), A-type AFM (A), C-type AFM (C), and G-type AFM (G) order parameters with ($x$, $y$, $z$) spin orientation; \textit{e.g.}, $C_y$ indicates the C-AFM order in which the spins are parallel along the global $y$ axis. 
It is shown that three magnetic order parameters belong to the same irreducible representation, \textit{i.e.}, $\{F_x, C_y, G_z\}$ orders belong to $m\Gamma_2$. 
It implies that the $C_y$ spin configuration is allowed to cause weakferromagnetic spin canting toward the $x$ direction and weak-antiferromagnetic spin canting with $G_z$ spin component without further symmetry reduction as being  
consistent with earlier symmetry analyses in Ref. \onlinecite{TrevesPRB1962, Bertaut1968_sym, Solovyev1996_lacro3}. 
This is also consistent with an experimental observation of the $C_y$ ground state with the $F_x$-type  weakferromagnetism  in Ref.~\onlinecite{Komarek2008_cacro3}. 
In addition, it must be noted that the magnetic symmetry allows the finite AHC in $C_y$ or $G_z$ order as well as in $F_x$ order even %\sout{without invoking the weakferromangnetic spin canting} 
if the net magnetization is negligible. 
This is because the physical properties of those three magnetic orders can be regarded equivalent from a symmetry point of view. 
This unusual symmetry property that AFM order and FM order share the same irreducible representation comes from  the nonsymmorphic symmetry operations.  For instance, $C_{2x}+(\frac{1}{2}\frac{1}{2}0)$ screw operation flips the $y$ and $z$  spin components located at (000) site and transfers it to $(\frac{1}{2}\frac{1}{2}0)$ site. The transfer between different spin sublattices in the C-AFM configuration makes $C_y$ order invariant under the screw operation as it makes $F_x$ order invariant as well. 
%KY note: Ofer wrote on existence of Fz instead of Fx. don't cite his paper here. 
%the  From ~\cref{table:symmetry}, there are four types of magnetic ordering, i.e., GxAyFz, AxGyCz, CxFyAz, and FxCyGz, associated with the four different IRs.
%All possible patterns (\textit{xyz} components) fit into one of four magnetic structures (FM and A-type, G-type, and C-type AFM) compatible with the \textit{Pbnm} space group. In this paper, we will represent each pattern by the major component with the largest projected spin moments, e.g., as C$_y$-type AFM state for C-type AFM configuration with the y axis spin moment. (KY: this is not true)
%\ky{I am here after Alberobello}
%In the nonsymmorphic space group, projections to some axes components are invariant under the translation transformation. (KY: I don't agree) For example, the FM configuration along the \textit{x}-axis and a C-type AFM pattern along the \textit{y}-axis are invariant under the $C^1_{2x}$ symmetry operation, same for the G-type AFM pattern along the \textit{z}-axis spin moment, expressed as a group $F_xC_yG_z$ ($m\Gamma2$).  For C-type AFM with spin aligns in \textit{y}-axis direction, it is allowed to have a ferromagnetic component along \textit{x} direction ($m_x$) and a non-zero Berry curvature $\Omega_{yz}$. 
%\ky{I bring the paragraph from introduction to here: 
The symmetry analysis here is also consistent with previous theoretical work on 
orthorhombic LaCrO$_3$, in which the finite FM orbital magnetization and the  optical nonreciprocity were predicted in the AFM configurations~\cite{Solovyev1996_lacro3}.  
%; it has been reported that AFM spin order coupled through the spin-orbit coupling (SOC) and the lattice distortion can lead to a nonvanishing ferromagnetic component of the orbital magnetization and optical conductivity in the LaCrO\3, owning to a similar structure but in insulating phase~\cite{Solovyev1996_lacro3}. 
%
%In the following section, we will focus on the magnetic symmetry and property of $C_y$-AFM since  is the magnetic ground state of \cacro. 
%\blue{

The magnetic space group %of C-type AFM with  S$||$010 
under $\{F_x, C_y, G_z\}$ order 
is type III \textit{Pbn'm'} (\textit{Pn'm'a} in the standard setting), 
%}%\ky{write also msg in standard setting with MSG ``type"} %(cf. ~\cref{table:symmetry}) 
containing four unitary operations \{$E$, $I$, $C_{2x}+(\frac{1}{2} \frac{1}{2} 0)$, $m_{x}+(\frac{1}{2} \frac{1}{2} 0)$\} and four antiunitary operations  \{$C_{2y}\theta+(\frac{1}{2} \frac{1}{2} \frac{1}{2})$, $C_{2z}\theta +(0 0 \frac{1}{2})$, $m_y\theta +(\frac{1}{2} \frac{1}{2} \frac{1}{2})$, $m_{z}\theta+(0 0 \frac{1}{2}) $\}, where $\theta$ denotes the antiunitary time-reversal operator.
%\ky{KY comes here.}

\begin{table*} [t]
\centering\def\arraystretch{1.1}
\begin{ruledtabular}
  \begin{tabular}{c c c c c c c c c} 
     IR of $Pbnm 1'$ & $E$ & $C_{2x}+(\frac{1}{2} \frac{1}{2} 0)$ & $C_{2y}+(\frac{1}{2} \frac{1}{2} \frac{1}{2})$ & $I$  & $\theta$  & Nonzero $M$ &Nonzero AHC & Magnetic \\
  & & & & & & component & component &  space group \\
  \hline
  $m\Gamma_1$: $A_x$, $C_z$, $G_y$ &1 & 1 & 1 & 1 & -1 & - & -& $Pbnm$ \\
  $m\Gamma_2$: $F_x$, $C_y$, $G_z$ & 1 & 1 & -1 & 1 & -1 & -$M_x$& $\sigma_{yz}$ & $Pbn'm'$ \\
  $m\Gamma_3$: $F_y$, $A_z$, $C_x$ & 1 & -1 & 1 & 1 & -1 & $M_y$& $\sigma_{zx}$ & $Pb'nm'$\\
  $m\Gamma_4$: $F_z$, $A_y$, $G_x$ & 1 & -1 & -1 & 1 & -1 & $M_z$& $\sigma_{xy}$ & $Pb'n'm$\\

  \end{tabular}
 \end{ruledtabular}
 \caption{ 
Transformation properties in nonmagnetic space group $Pbnm 1'$ for FM ($F_{\alpha}$) and AFM ($A_{\alpha}, C_{\alpha},$ and $G_{\alpha}$) magnetic ordering parameters with $\alpha = x,  y, z$ components in the global frame. %\blue{KY found that there are local frame and global frame with same $x, y, z$ labels in this paper!} 
Only the generators of symmetry operations are shown. The group
elements denote the identity, two screws, inversion, and the time reversal. Irreducible representation (IR)'s name is taken 
from ISODISTORT software~\cite{isodistort}.%\ky{In English, $\times$ symbol does not have a negative meaning. Often it has positive meaning. you should put - instead. For $m\Gamma_1$, make 1 be subscript. }
\label{table:symmetry}}
\end{table*}

\begin{table} [t]
\centering\def\arraystretch{1.1}
\begin{ruledtabular}
  \begin{tabular}{c c c} 
  & \{$k_x, k_y, k_z$\} & \{$\Omega_x, \Omega_y, \Omega_z$\} \\
  \hline
  $E$   & \{$k_x, k_y, k_z$\} & \{$\Omega_x, \Omega_y, \Omega_z$\}  \\
  $I$   & \{$-k_x, -k_y, -k_z$\} & \{$\Omega_x, \Omega_y, \Omega_z$\}  \\
  $C_{2x}$ & \{$k_x, -k_y, -k_z$\} & \{$\Omega_x, -\Omega_y, -\Omega_z$\}\\
  $m_x$   & \{$-k_x, k_y, k_z$\} & \{$\Omega_x, -\Omega_y, -\Omega_z$\}  \\
  $C_{2y}\theta$ & \{$-k_x, k_y, -k_z$\} & \{$\Omega_x, -\Omega_y, \Omega_z$\}  \\
  $C_{2z}\theta$ & \{$-k_x, -k_y, k_z$\} & \{$\Omega_x, \Omega_y, -\Omega_z$\}  \\
  $m_y\theta$   & \{$k_x, -k_y, k_z$\} & \{$\Omega_x, -\Omega_y, \Omega_z$\} \\
  $m_z\theta$   & \{$k_x, k_y, -k_z$\} & \{$\Omega_x, \Omega_y, -\Omega_z$\}\\
  \end{tabular}
 \end{ruledtabular}
 \caption{ 
Transformation rules for crystal momentum $\bm{k}$ and %\sout{spin-1/2 operators} 
Berry curvature $\bm{\Omega}$ under symmetry operations in $Pbn'm'$ magnetic space group under $\{F_x, C_y, G_z\}$ order. The translation operations in the spiral and glide symmetry operations were removed in the $k$ space.  %Note that we neglect the translation operations for simplicity.\ky{This is not for simplicity. This is because there is no translation in k-space!}
%\ky{replace s by $\Omega$!!} 
%in this table.  %\thao{add Omega xyz follow Suzuki's paper [PHYSICAL REVIEW B 95, 094406 (2017)]}
\label{table:Pbn'm'}}
\end{table}

\subsection{\label{ssec:ahc}Anomalous Hall conductivity} 
  
The AHC was calculated as integrating  the Berry curvatures with a summation over the occupied states in the Brillouin zone by WANNIER90 code~\cite{WangVanderbilt2006, wannier90}:%~\cite{WangVanderbilt2006}  
\begin{eqnarray}
    \sigma_{\alpha\beta}=-\frac{e^2}{\hslash} \int_{\rm BZ} \frac{d{\bm k}}{(2\pi)^3}\sum_n f_n({\bm k}) \Omega_{n,\alpha\beta}({\bm k}),
\end{eqnarray}
%\ky{cite the reference} 
where $n$ is the band index,  $\alpha$ and $\beta$ %\sout{respects to} 
are the global Cartesian directions $(x,y,z)$; $\alpha \ne \beta$  for the AHC components and  $f_n({\bm k})$ is the occupation factors at the $\bm{k}$ point. %\sout{determined from the difference between the eigenvalue of the Bloch states $\epsilon_{n{\bm k}}$ and the Fermi energy}
%\ky{Cite the papers from which you took the equations. I don't understand your definition and don't find connection to the following equation.}. The %band-resolved 
The Berry curvature was calculated  %written as an antisymmetric tensor
by 
%\thao{Eq. below taken from user_guide_wannier90_ver3.1}

\begin{eqnarray}
    \Omega_{n,\alpha\beta}({\bm k})=-2{\rm Im}\left< \nabla_{k_\alpha}u_{n{\bm k}}|\nabla_{k_\beta}u_{n{\bm k}} \right> 
\end{eqnarray} 
in Wannier90 code~\cite{LVTS12, wannier90}. 
%The derivative $\nabla_{k_\alpha}u_{n{\bm k}}$ can be obtained 
Here, $u_{n{\bm k}}$ are the cell-periodic Bloch functions for $n$-th band, projected onto Wannier functions $\left| {\bm R} n \right>$ by 
\begin{eqnarray}
u_{n{\bm k}} = \sum_{\bm R} e^{-i{\bm k}\cdot({\bm r}-{\bm R})}\left| {\bm R} n \right>. 
\end{eqnarray} 

By applying the conventional perturbation theory, the Berry curvature can be cast into the form of a Kubo-like formula, 

\begin{eqnarray}
 %   \Omega_{n,\alpha\beta}({\bm k})=-2\hslash^2 {\rm Im} \sum_{m\ne n} \frac{\bar{H}_{nm,\alpha}({\bm k})\bar{H}_{mn,\beta}({\bm k})}{[\epsilon_m({\bm k})-\epsilon_n({\bm k})]^2}, 
 \label{eqn:kubo}
    \Omega_{n,\alpha\beta}({\bm k})=-2\hslash^2 {\rm Im} \sum_{m\ne n} \frac{v_{nm,\alpha}({\bm k})v_{mn,\beta}({\bm k})}{[\epsilon_m({\bm k})-\epsilon_n({\bm k})]^2}, 
\end{eqnarray}

%The Berry curvature was computed by following Kubo-like formula  %with a Wannier-based approach using 
%\ky{there is no such a word as ``Wannier-based approach''} \thao{written in Wang2006}
%\thao{This is Kubo fomular and WANNIER90 does not use this EQ.}
%\begin{eqnarray}
%    \Omega_{n,\alpha\beta}({\bm k})=-2\hslash^2 {\rm Im} \sum_{m\ne n} \frac{v_{nm,\alpha}({\bm k})v_{mn,\beta}({\bm k})}{[\epsilon_m({\bm k})-\epsilon_n({\bm k})]^2}, 
%\end{eqnarray}
%\ky{bring Kubo formula back here.}

where $\epsilon_n({\bm k})$ is the eigenenergy for $n$-th band at a given ${\bm k}$ point and %\orange{$\bar{H}_{nm,\alpha}({\bm k})$ is the approximate tight-binding velocity operator between a pair of the occupied bands $n$ and $m$ obtained within the Wannier representation.~\cite{WangVanderbilt2006}}
$v_{nm,\alpha}({\bm k})$ is the matrix element of the velocity operator between the occupied $n$ state and the unoccupied $m$ state.
%$\hat{v}_\alpha=$ \ky{??} along $\alpha$ direction is given by
%\begin{eqnarray}
%    v_{nm,\alpha}({\bm k})=\left< \psi_{n{\bm k}} |\hat{v}_\alpha| \psi_{n{\bm k}}\right> = \frac{1}{\hslash} \left< u_{n{\bm k}} \left|\frac{\partial \hat{H}({\bm k})}{\partial k_\alpha}\right| u_{m{\bm k}}\right>
%\end{eqnarray}
%
%with $\hat{H}({\bm k})=e^{-i{\bf k}\cdot{\bf r}} \hat{H} e^{i{\bf k}\cdot{\bf r}}$ and $u_{m{\bm k}}$ are the cell-periodic Bloch functions.
%
%
The AHC and Berry curvature can be regarded as %\sout{pseudovectors} 
axial vectors like the spin momentum under the magnetic symmetry operations %\sout{with their tensor components}
in their vector form: ${\bm \sigma}=(\sigma_{x}, \sigma_{y}, \sigma_{z}) = (\sigma_{yz}, \sigma_{zx}, \sigma_{xy})$ and 
${\bm \Omega}=(\Omega_{x}, \Omega_{y}, \Omega_{z})=(\Omega_{yz}, \Omega_{zx}, \Omega_{xy})$,  
where  $n$- and $k$-dependency was omitted for simplicity. 
%\ky{You wrote Omega as a function of k. Now you omit k. That is non-consistent. Use the same variable throughout the paper!!}
%\ky{Explain how they are transformed under symmetry operations such as time-reversal operation.}
%\blue{[Rephrase the following text taken from Gurung-Tsymbal paper with proper citations] 
%\blue{
%For example, 
Since $\Omega$ is odd with respect to time-reversal symmetry flipping the $k$ vector, %i.e. $\Omega(-k) = −\Omega(k)$,
the summation of $\Omega$ and accordingly the AHC are zero for non-magnetic materials. Similarly, the symmetry operations transforming $k$ to $-k$ and $\Omega$ to $-\Omega$
simultaneously, the AHC vanishes. 
~\cref{table:Pbn'm'} shows the transformation of $k$ and $\Omega$ under $\{F_x, C_y, G_z\}$ order; here only summation of $\Omega_{x}$  in $k$ space (=$\sigma_{x}$) can be non-zero whereas $\Omega_{y}$ and $\Omega_{z}$ cancel out in summation. 
 %One point we emphasize is that 
 All the components of AHC vanish if SOC is %\sout{ignored} 
not considered in the calculations. % owing to the effective $R_sT$
In the absence of SOC, the spin direction does not affect the orbital nor charge state and hence %\sout{the C-AFM order has the common symmetric property of $C_x$, $C_y$, and $C_z$-AFM orders, $i.e.,$ invariant under all screws and glides operations of $Pbnm$ space group,} \red{
the spin state is transformed under symmetry operations defined in a black-and-white group (see Appendix A), 
preventing the spin polarization and AHC in any direction. 
%symmetry.\cite{Suzuki2017PRB_multipole} %\sout{Without SOC, the AHC is zero} %\ky{don't repeat the sentence. Readers are not stupid.} 
%since the non-zero Berry curvature emerging at non-symmetric $k$ points  cancel out in integration. 
%as a result of the effective {\it PT} symmetry (KY: PT is violated due to Zunger).\cite{Gosalbez-Martinez-VanderbiltPRB2015}  
%Without SOC, the Berry curvature vanishes \ky{is it true?} identically as a result of the effective {\it PT} symmetry as in FM case.\cite{Gosalbez-Martinez-VanderbiltPRB2015}
%}
%}

\begin{figure}
    \centering
    \includegraphics[width=0.9\columnwidth]{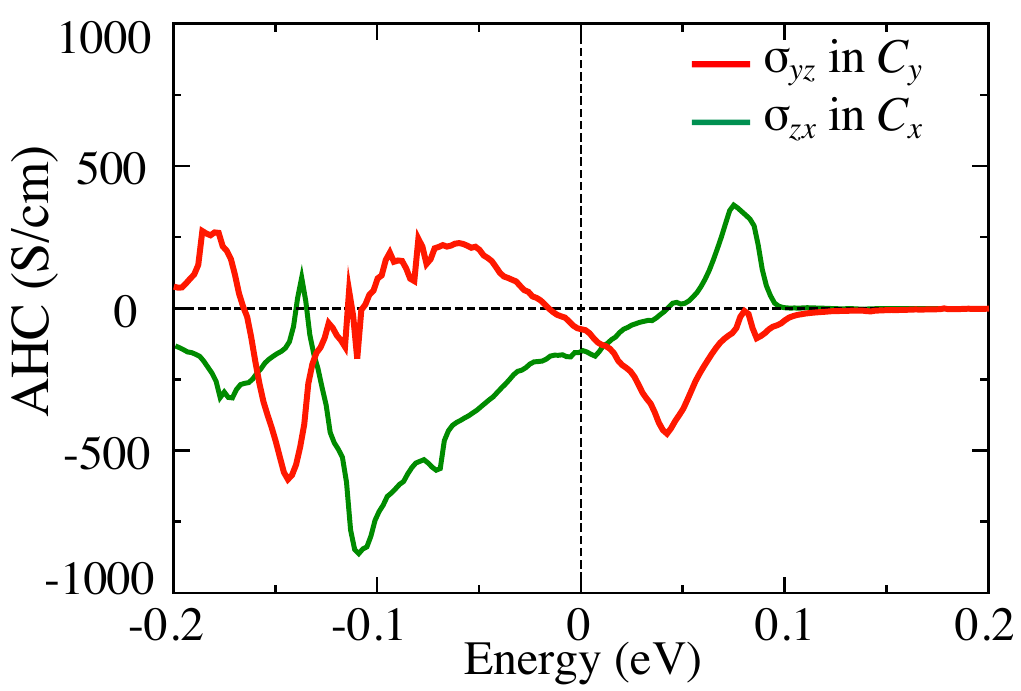}
    \caption{ %\thao{remove Cy and Cx. modify the legend to sigmayz in Cy, sigmazx in Cx. sigmaAHC should be sigma or AHC only, make lager font for Energy} %   \thao{Make Cy in RED line and thicker} 
    Calculated AHC tensor components; %(orange curve) 
    $\sigma_{yz}$ in $C_y$-AFM  configuration and %(green curve) 
    $\sigma_{zx}$ in $C_x$-AFM configuration as  functions of energy. Fermi energy is set to zero.}
    \label{fig:ahc} 
\end{figure}

%\thao{Equations and definitions are taken from HuyenPRB2019, Zhou2019 and Wang2006(Wannier90 AHC calculation)}
Figure~\ref{fig:ahc} shows the calculated AHC in C-AFM configuration. All the tensor components 
%in $C_x$, $C_y$, and $C_z$-AFM configurations
were found to be negligible except $\sigma_{yz}$ in $C_y$ and $\sigma_{zx}$ in $C_x$ AFM configurations being in agreement with our symmetry analysis (cf. \cref{table:symmetry}). %\sout{Indeed, $C_x$- and $C_y$-AFM orders has same symmetrical property as $F_x$ and $F_y$-orders, respectively, while  $C_z$-AFM order does not belong to IR of any FM order, resulting in zero AHC.} \ky{Why do you repeat the phrase here?} 
The values of $\sigma_{yz}$ and $\sigma_{zx}$ at Fermi energy were calculated as -74 S/cm and -149 S/cm, respectively.\footnote{We obtained larger AHC value in FM order;  $\sigma_{yz}=$ -150 S/cm with  $F_x$ order in CaCrO$_3$.}  
 % collinear $C_y$-AFM CaCrO$_3$ i
These values are comparable to that in  noncollinear AFM Mn\3Sn, reported as  
%$\sigma_{{\rm cubic}(111)}=$ 218 S/cm in Mn\3Ir~\cite{Chen2014}. 
$\sigma_{yz}^{\rm DFT}=$ 129 S/cm~\cite{Suzuki2017PRB_multipole}  ($\sigma_{yz}^{\rm EXP}=$ 100 S/cm~\cite{Nakatsuji2015_mn3sn}).  
Despite the sizable AHC values, 
CaCrO$_3$ yields only tiny weakferromagnetic spin canting with the net magnetization   $M_x$ = 0.03 $\mu_{\rm B}$/f.u.  obtained after optimizing the spin directions in $C_y$-AFM configuration. 
This clearly indicates that the conductive electrons sensitively experience the Berry curvature as a fictitious magnetic field in $k$ space instead of the magnetic field caused by the spontaneous magnetization in real space. 
%In CaCrO$_3$,  
%\sout{Even though the value of AHC is not as large as the one in the ferromagnetic case, our finding of nonzero AHC on collinear AFM paves the way for the new discovery of AHC.} %\ky{so you want to say the AFC in C-AFM is small? Then we need to correct the abstract in that way.}
%However, this value is still one order smaller than the famous ferromagnetic bcc iron (~750S/cm)\cite{Yao2004, WangVanderbilt2006}.

The non-monotonous behavior of  $\sigma_{yz}$ and $\sigma_{zx}$ with respect to the energy is closely associated with the multi-fold $t_{2g}$ band structure around Fermi energy, showing  a different trend from the result calculated by 
%consistent with the symmetry analysis and the prediction by 
a tight-binding model showing simpler energy dependency~\cite{Naka2022_AHC}. 
%This emphasizes the advantage of the combination of DFT and Wannier function approaches. 
%\blue{[Maybe we can rephrase the text from WYSV paper]: 
%In particular, % the interpolated band structure is able to reproduce 
In fact, by virtue of the Wannier interpolation, 
we can quantitatively evaluate the magnitude of AHC that is sensitive to tiny features of the band structure, such as spin-orbit-induced band anti-crossing in a tiny $k$ space volume. 
The AHC values can be enhanced by shifting the Fermi energy either upward or downward (see ~\cref{fig:ahc}). For example, $\sigma_{yz}$ = -441 S/cm at $E=E_{\rm F}$ + 36 meV in $C_y$-AFM and   %\ky{what are you writing here? Do you want to discuss the enhancement of $\sigma_yz$ at EF + 400meV?? } \thao{It's the peak at EF+0.04eV(36meV) of red curve}
% the value is two times larger than the calculated AHC in by Chen2014 noncollinear AFM Mn\3Ir.~\cite{Chen2014}
%experimental AHC in Mn\3Ge 130 S/cm.~\cite{Chen_Nakatsuji2021} %calculated AHC for Mn3Ir is 218 S/cm by Chen2014
%The large values of $\sigma_{yz}$ (-601 S/cm) was found at 144meV 
%For $C_x$-AFM configuration, two large peaks of %$\sigma_{zx}$ = 109 meV below the Fermi energy  (-862 S/cm) 
$\sigma_{zx}$ = 363 S/cm at $E=E_{\rm F}$ + 75 meV in   $C_x$-AFM configurations. 
These values are comparable to the AHC value ($\sigma_{xy} \sim $ 750 S/cm) calculated in FM bcc Fe,~\cite{Yao2004, WangVanderbilt2006} %\ky{just to check: your unit S/cm is same as $(\Omega cm) ^{-1}$  in WYSV paper?}\thao{YES, IT IS}  
%The change of magnitude of AHC values as shifting the Fermi energy implies 
implying the possible  enhancement of AHC in CaCrO$_3$ via chemical doping in practice.  %\sout{for further spintronics applications}.%\ky{What application do you consider?}

\begin{figure}
    \centering
\includegraphics[width=0.9\columnwidth]{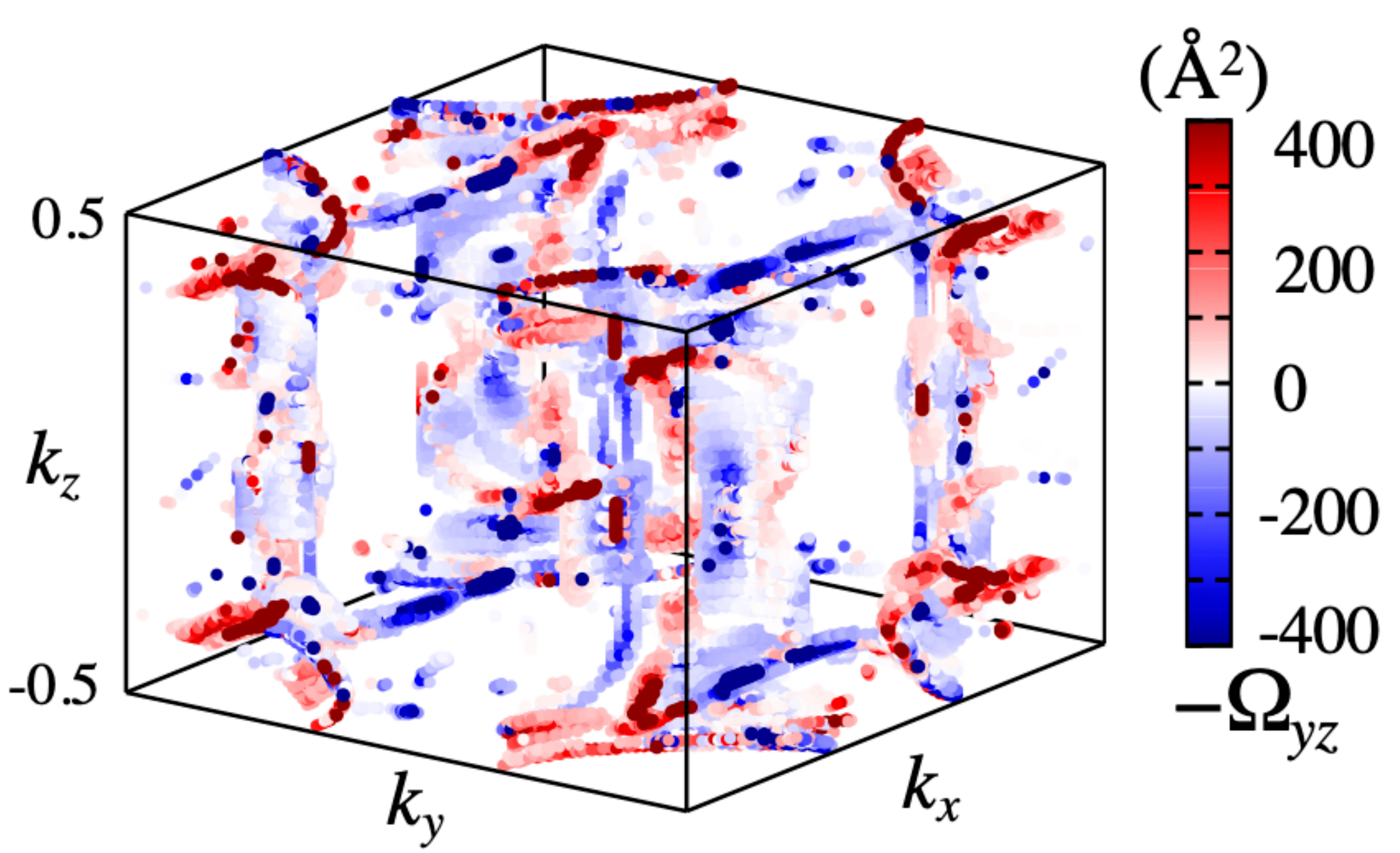}
    \caption{  %\sout{Hot spot of Berry curvature} \ky{Don't use such ill-defined terms. Check how other people explain the property!} 
   %\ky{Delete the layers and shaded part.} 
     Distribution of Berry curvatures $-\Omega_{yz}$ ({\AA$^2$}) over the octahedral Brillouin zone in $C_y$-AFM configuration. 
%(b-c) Berry curvatures and (d-e) nodal lines (spin-splitting bands with tiny band gap at Fermi energy) in the $k_z=0.50$ and $k_z=0.42$ planes, respectively.\thao{correct kz=0.42} 
}
    \label{fig:berry} 
\end{figure}

\subsection{\label{ssec:mechanism}Berry curvature and band splitting}

%\ky{ will see this section after Nov.28th}
%\sout{Large Berry curvature and a huge AHC in AFM materials are usually caused by the band structure near the Fermi level, such as Weyl nodes or gapped nodal lines~\cite{Chen_Nakatsuji2021, Huyen2021, Zhao2021}. 
%In this session, we point out the origin of finite AHC by looking at the hot spot of Berry curvature in the BZ.}\ky{It was written in Introduction.} 

%\thao{Question to solve: 1) How symmetry involve in the cancellation of Berry curvature in y and z directions? 2) What is the behavior of Sx in k-space and in real space. Note that Cy is an order parameter, the magnetic moment can be written as Cy=S1y-S2y+S3y-S4y, Sx and Sz (spin) components are zero in Cy order, why it show spin polarization for Sx in the band structure while the magnetic moment along x is almost zero (0.03muB)}

In the following, to clarify the microscopic origin of the sizable AHC, 
we focus on the $C_y$-AFM configuration. 
Figure \ref{fig:berry} shows the distribution of the Berry curvatures $\Omega_{yz}$ in the orthorhombic Brillouin zone. %We note that even though positive and negative values of Berry curvature might be canceled out %\ky{to cancel out is an intransitive verb!} in the \textbf{k}-space, the hot spot area is still expected to have greatly contributed to the AHC. 
It can be seen that the shape formed by high Berry curvature spots resembles some Fermi surfaces (shown in ~\cref{fig:fermisurface} (b-e)) manifesting that the Berry curvatures stem mainly from the topologically nontrivial points in the vicinity of the Fermi surfaces~\cite{WangVanderbilt2007, Haldane2004}.

\begin{figure}
    \centering
    \includegraphics[width=0.9\columnwidth]{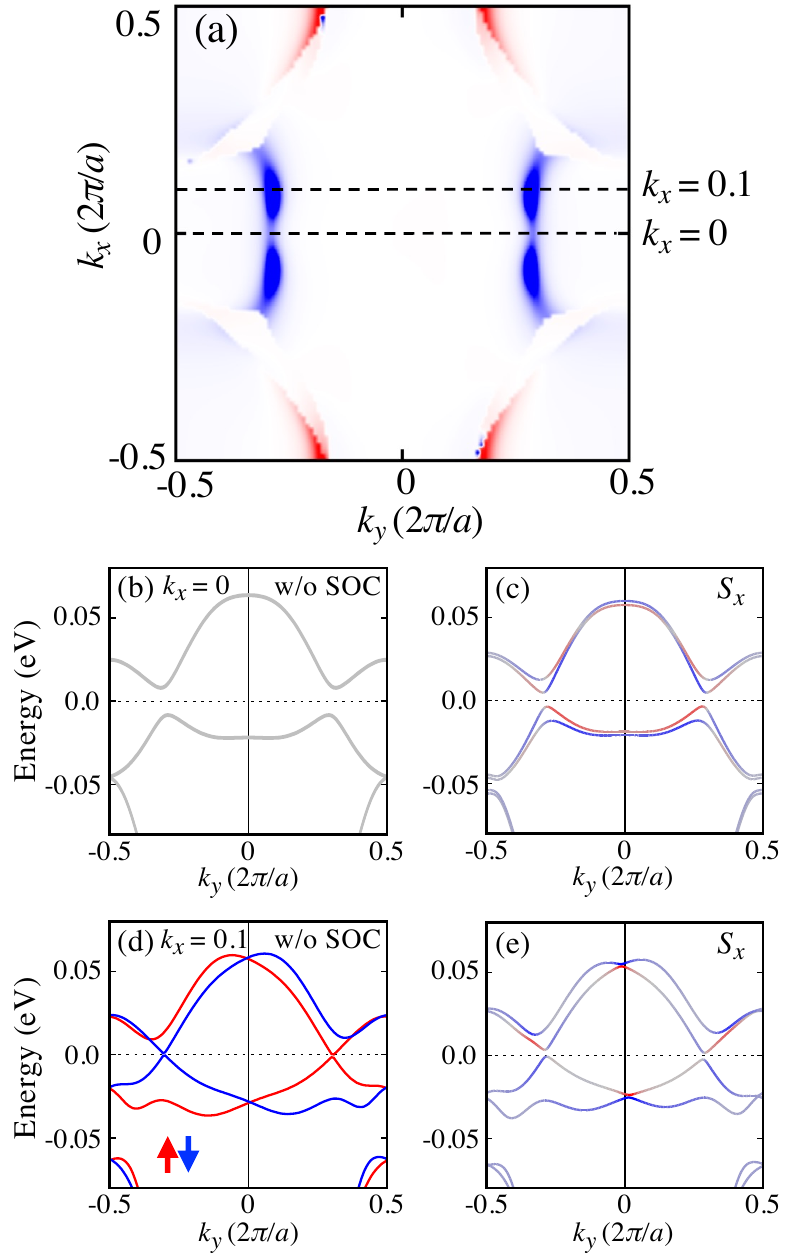}
    \caption{ %\thao{ make no SOC band in larger energy range, then add band sticking effect to that band}
    %(a)Schematic pictures of spin-split band structure without and with SOC for (top) FM and (bottom) AFM orders.
    %\ky{Remove (a) and (b) panels. Make the solid line dashed in (d) and make the font size bigger (subscript x and y are too small now). Make the y-axis larger (change the aspect ratio) in (e)-(h) to show the anticrossing clearer. } 
    %(a) Gapped nodal lines on $k_z=0.5$ plane. \ky{Explain what you plot exactly. I think you plot the spot where the two band energies are closer than 0.x meV just near the Fermi energy.} 
    %\ky{change the aspect ratio of lower band structure plots so that they look like regular squares.} 
    (a) Color map of Berry curvature $\Omega_{yz}$ in  $k_z=0.42$ plane under  $C_y$-AFM configuration. The color range  is set same as in  ~\cref{fig:berry}.
    Band structure along the $k_y$ direction corresponding to two dashed lines in (a) in the same plane;   (b) and (c) for $k_x=0$  without SOC and  with SOC, respectively; (d) and (e)  for $k_x=0.1$ without SOC and  with SOC, respectively. In (c) and (e),  
     the spin polarization along the $x$ direction ($S_x$) is shown by red and blue colors. In (d), the spin polarization does not have any preferential direction.  %\thao{adjust the color in panel (e) range -0.15:0.15, in panel (c) range -0.3:0.3.}
    %\ky{I don't like the labels touching graph plots. Make more space or shift labels to avoid them. I want the lower graphs in regular square aspect ratio. Elongate them a bit more. }
%   \ky{Put "w/ SOC" label in the right figures. }\thao{Updated. Check and delete this comment} %I don't know whether the spin arrows in panel (d) are appropriate or not. In this case, spin is not an arrow but just a state. Let's remove the arrows and write "red and blue colors indicate different spin states." in the caption.}
    }
    \label{fig:spin_polarized_bands} 
\end{figure}

%\sout{In fact}\ky{stop to use "in fact" many times}, the expression of the Berry curvature regarded as a Fermi-surface property has been examined by Haldane's and other works.~\cite{WangVanderbilt2007, Haldane2004}.\ky{What is the point?What do they discuss? }
%
%\ky{Reconstruct the following paragraphs in more readable order. 1) explain the Berry curvature slices with the symmetry, 2) focus on the nodal lines at kz=0.4, 3) explain the spin-splitting phenomena.} 
%\ky{here the figure of $\Omega$ slices and symmetry property should be explained} 
Significantly high Berry curvatures %to the AHC %in the momentum space %\ky{contribution to what? don't use this word by itself.}
were detected at $k_z=0.42$  plane in the Brillouin zone.    %. The significant %\sout{contributions} $are 
As shown in 
~\cref{fig:spin_polarized_bands} (a), the high Berry curvatures plot visualizes %a circular curve around the BZ corner and  
the hot spots, {\it i.e.}, two broad lines running parallel to the $k_x$ axis. These spots are located along the nodal lines opening a small gap near the Fermi energy (cf. Fig.~\ref{fig:spin_polarized_bands} (c) and (e)) 
and originating from the inter-band interaction between the 11th and 12th bands (also see the corresponding Fermi surfaces at Fig.~\ref{fig:fermisurface} (b)).  

Although one may think that those two bands ({\it i.e.,} up- and down-spin polarized bands coming from up- and down-spin Cr sites, respectively) were degenerate in the collinear AFM order, the magnetic symmetry in fact allows the spin splitting {\it even without SOC} at generic $k$ points owing to the effective $PT$ violation;  CaCrO$_3$ can be categorized into SST-4A type like LaMnO$_3$ in Ref. \onlinecite{Yuan_Zunger2021}. The spin degeneracy is protected by some symmetry that couples the up-spin and down-spin sites at the higher symmetric $k$ points along the $\Lambda$H line ($k_x$=0.0; $-0.5<k_y<0.5$; $k_z$=0.42) as shown in Fig.~\ref{fig:spin_polarized_bands} (b) (for the detail symmetry analysis, see Appendix A). On the other hand, at non-symmetric $k$ points ($k_x$=0.1; $-0.5<k_y<0.5$; $k_z$=0.42) shown in Fig. \ref{fig:spin_polarized_bands} (d), the spin degeneracy is lifted even without SOC.   
%\sout{Figure \ref{fig:spin_polarized_bands} (d) indeed shows that the spin degeneracy is lifted at nonsymmetric $k$ points ($k_x$=0.1; $-0.5<k_y<0.5$; $k_z$=0.42) without SOC.}
%SOC lifts the spin degeneracy  except along the nodal lines. 
A couple of the spin-polarized bands are crossing exactly at the Fermi energy, resulting in the nodal lines. 
When SOC is turned on, 
a small gap appears by anticrossing effect and the two bands  hybridize across the gap through the SOC interaction.  
This in turn lets the bands originally having $S_y$ polarization acquire weak $S_x$ polarization as shown in Fig.~\ref{fig:spin_polarized_bands} (e). Simultaneously, $\Omega_{yz}$ manifests itself and shows strong enhancement due to the tiny anti-crossing gap along the nodal lines due to Eq.~\ref{eqn:kubo}. 
In terms of magnetic symmetry, $m_x$ and $m_y\theta$ symmetry keeps the $\Omega_{yz}$ invariant and makes the hot-spot shape symmetric along the $k_x$ and $k_y$ direction in the plane. 
%this makes spin-splitting  spin polarization allowing the hybridization of the states from opposite spins. 

%\thao{~\cref{fig:spin_polarized_bands}(b) and (c) were not mentioned.} -> KY: Sorry. 
% except along the nodal lines). 
 %Note that the gapped nodal line, in this case, is defined as the gap between two bands (split by SOC) is less than 0.01eV. %\ky{Is it the correct definition of gapped nodal line? check literature. }
%In most cases, t
%While the {\it `gapped nodal line'} is normally defined as the \sout{gapless} nodal line that gaps out with the introduction of SOC \sout{according to the magnetization direction},~\cite{Manna_Reviews2018, Shukla_PRB2021}, the situation is slightly different in our case.
%Two-fold degeneracy bands are \red{accidentally closed to each other (but not crossing)} and }

%Band splitting by SOC occurs everywhere in the whole BZ, however, the spin degeneracy at the edge \ky{what edge?} is protected by the symmetry. The band structure along the edge in ~\cref{fig:band_splitting} (a-c) implies that the splitting at 0.1eV below the Fermi energy \ky{this expression of energy is not correct} and the mixture of spin-up and spin-down states due to the symmetry (in this case, $\sigma_x'$, $\sigma_z$, $C_{2y}’$) result in a zero Berry curvature \ky{why does mixture of spin states result in zero curvature???}. In contrast, by moving down to $k_z=0.42$ plane in ~\cref{fig:band_splitting} (d-f), the bands are no longer sticking together, bringing the crossing bands to the Fermi energy. The symmetry reduction at general points allows spin-up and spin-down splitting when the spin aligns along the $S_y$ direction. %\ky{Can you plot Fig.5 without SOC? Are you sure that there is no spin splitting without SOC?}\thao{Along the k-path plotted in Fig5, I am sure there is no spin splitting without SOC, I have slide for it we can check together} 
%\thao{Delete this paragraph}
%\sout{Last, let us comment on the contribution to the AHC in real space via the hopping parameter. The ferromagnetic contributions come from hoppings between the half-occupied $t_{2g}$ orbitals via oxygen. In particular, we found the large hopping between bonding and antibonding $d_{xz-yz}$ and $d_{xz+yz}$ orbitals. These are here rather large due to a very strong $p-d$ hybridization due to substantial GdFeO\3-type distortion.} %\thao{Should we mention imaginary hopping? Must check!}

\subsection{\label{ssec:distortion}Effect of structural distortion on anomalous Hall conductivity}

%\orange{The GdFeO\3-type distortion is supposed to play a crucial role in the appearance of AHC in orthorhombic perovskite. In Ref.~\onlinecite{Naka2022_AHC}, Naka \etal\ have discussed the essential of distortion to AHC in term of the orbital mixing effect together with the SOC, in which, the nonzero Berry curvature is activated by the interorbital matrix elements of the effective hopping between d orbitals through the p orbitals, owing to the GdFeO\3-type distortion. }
 
%\thao{Parti-Sci.rep.2018:Due to the significant tilting and rotation of oxygen octahedra, the system has the orthorhombic perovskite crystal structure with four Ir sublattices and Pbnm nonsymmorphic space group~\cite{Patri_Sci.rep2018}}
%\ky{Explain why structural distortion is important: explain what Naka san claims in his paper }
%\ky{and explain that octahedron tilting is related to the non-symmorphic symmetry} 

%\blue{

As discussed in the preceding section, the non-symmorphic symmetry operation that relates the opposite spin sites in AFM order is a key factor to link the AFM order to FM order and  to give rise to the AHE in collinear AFM materials. 
The orthorhombic crystal structure of CaCrO$_3$ accommodates a three-dimensional  rotation of the octahedra, known as the GdFeO$_3$ rotation, concurrently with a Jahn-Teller distortion with respect to the cubic perovskite structure. 
Considering that the cubic perovskite structure shows symmorphic $Pm\bar{3}m$ space group, %and it does not activate AHE in AFM orders, 
we can deduce that the GdFeO$_3$ rotation drives the AHE and the resulted AHC increases with the distortion. 

In order to examine the effect, %of the  structural distortion upon AHE, 
we performed additional DFT calculations of AHC in MgCrO$_3$ and SrCrO$_3$, replacing A-site cation in CaCrO$_3$ by smaller and larger elements, respectively. 
The crystal structures were relaxed starting from \cacro structure.
%Let us draw attention to the relation between octahedral tilting (i.e. GdFeO\3-like tilting) and the size of AHC. 
Table~\ref{tabl:lattice} summarizes  the detail of the relaxed structure and 
%comparison of Cr-O-Cr bond angle and 
the calculated AHC in MgCrO\3, CaCrO\3, and SrCrO\3.   
The calculated Cr-O-Cr bond angles show that the GdFeO\3-type octahedral tilting is enhanced in MgCrO\3 and absent in SrCrO\3. 
This is consistent with an experimental observation that SrCrO\3 crystallizes in tetragonal $P4/mmm$ structure~\cite{komarek2011_srcro3} while MgCrO\3 structure has not been reported to our best knowledge. 
The resulted AHC shows that $\sigma_{xy}$ is zero for SrCrO\3 and $\sigma_{xy}$ for MgCrO\3 is smaller than that for CaCrO\3; the latter is 
counter-intuitive. 
In fact, the magnitude of AHC strongly depends on the detail of the band structure near the Fermi energy and therefore it is not directly controlled by tuning the structural distortion. Nevertheless, our result highlights the importance of structural distortion as a driving force of AHE. 

\begin{table}[hbt!]
%\centering\def\arraystretch{1.1}
\resizebox{\columnwidth}{!}{%
\begin{ruledtabular}
 \begin{tabular}{c c c c c c c c} 
  & $a$ & $b$ & $c$ & $\theta$ & $d_{\rm long}$ & $d_{\rm short}$ & $\sigma_{yz}$\\
  \hline
  MgCrO\3 & 4.9833 & 5.1886 & 7.3030 & 139.9 & 1.94 & 1.91 & 73.5\\  %\cacro(exp)~\cite{GOODENOUGH1968_cacro3} & 5.2870 & 5.3160 & 7.4860  & 155.21 & & \\
  \cacro & 5.2873 & 5.3566 & 7.4985 & 155.2 & 1.92 & 1.91 & 74.8\\
  %SrCrO\3 & 5.4241 & 5.4248 & 7.6639 & 179.5 & 1.91 & 1.91 & 0\\
  SrCrO\3 & 5.4245 & 5.4245 & 7.6639 & 180.0 & 1.91 & 1.91 & 0.0 \\
  \end{tabular}
 \end{ruledtabular}
 }
 \caption{ Optimized lattice constants $a$, $b$, and $c$ ({\AA}), Cr-O-Cr bond angle in CrO$_2$ plane $\theta$ ($^{\circ}$),  
 long and short bond distances between Cr and O atoms $d_{\rm long}$ ({\AA}) and $d_{\rm short}$ ({\AA}),  and  calculated AHC $\sigma_{yz}$ (S/cm)  at Fermi energy in $A$CrO$_3$ for $A$=Mg, Ca, and Sr under $C_y$-AFM order. 
 %\thao{Reduce to two decimal digits?}\ky{add space group, add middle bond length, explain which bond is long and short. }
}
\label{tabl:lattice}
\end{table}

\section{Summary} \label{sec:conclude}

%By means of first-principles calculations, we evaluated the anomalous Hall effect in collinear AFM \cacro. In particular, we found that the AHC in the collinear AFM phase is finite and as large as that in the FM case. The AHC is owing to the non-symmorphic symmetry that binds the AFM and FM order in the same IR. SOC gives rise to the spin-splitting and band anti-crossing yielding large Berry curvatures. 
The AHE in CaCrO$_3$ was predicted by means of first-principles calculation supported by symmetry analysis. 
The AHC was found to be sizable in the collinear C-AFM as the magnetic ground state and  
we revealed two essential roles of the non-symmorphic symmetry.    
(i) The screw and glide symmetry operations bind the AFM and FM order parameters in the same irreducible representation so that AHE is active in the AFM order.  (ii) The band-sticking effect at the Brillouin-zone surface makes Cr-$t_{2g}$ state form the narrow bands near the Fermi energy; a couple of those bands cause anticrossing and enhancement of the Berry curvature. 
%We hope that future experiments can verify our predictions of this unusual AHE in collinear antiferromagnets. 
We hope that our prediction of AHE in CaCrO$_3$ will be verified by future experiments. 
We also expect that our study will  provide an important step forward in the understanding of this unusual AHE and  in the exploration of related phenomena further in transition-metal oxides with their wide variety of structures and tunable magnetic properties. 
%exotic, perhaps unique, superconductor.

%This finding provides a guiding principle to explore further AHC in AFMs. 
%We analyzed the hotspots in the k-space that shows large Berry curvatures and located them along nodal lines at generic points, generating a large AHC with lifting degeneracy by SOC. 
%\ky{the conclusion is very boring. write the sales point of your study. I would say that AHC in AFM discussed here is owing to the non-symmorphic symmetry which bind the AFM and FM order in the same IR and causes the multiple degeneracy at BZ surfaces. On top of it, SOC gives rise to the spin-splitting and band anti-crossing yielding the large Berry curvatures. This finding provides a guiding principle to explore further AHC in AFMs.  }

\begin{acknowledgments}
We are grateful to M. Naka, M.-T. Suzuki and S. Picozzi for the fruitful discussions. 
This work was supported  by JST-CREST (Grant No. JPMJCR18T1). The computation in this work has been done using the facilities of the Supercomputer Center, the Institute for Solid State Physics, the University of Tokyo and Supercomputing System MASAMUNE-IMR in the Center for Computational Materials Science, Institute for Materials Research, Tohoku University (Project No. 20K0045). The crystallographic figure was generated using the VESTA program~\cite{vesta3}.
\end{acknowledgments}

\appendix

\crefalias{section}{appsec} 

%----------------------------------------
\section{Symmetry analysis for spin splitting   }
\label{sec:appendix}
%---------------------------------------

\begin{figure}[ht]
    \centering
\includegraphics[width=0.9\columnwidth]{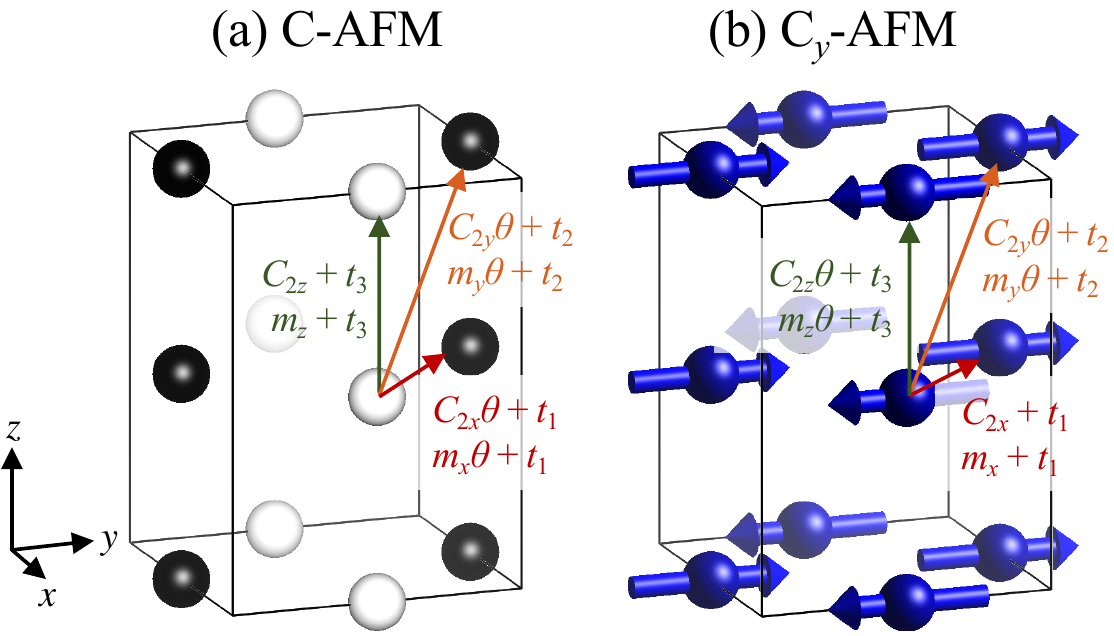}
    \caption{ Screw and glide symmetry operations in (a) C-AFM configuration with black-and-white  $Pb'n'm$  space group  %with $Pnma$ (?) black-and-white group 
    and (b) $C_y$-AFM configuration with magnetic $Pbn'm'$ space group.  %with $Pbn'm' (?)$ magnetic space group. 
 %      C-AFM order without SOC follows black-and-white  $Pb'n'm$  space group and $C_y$-AFM order with SOC follows magnetic $Pbn'm'$ space group. 
 The translation vectors are shown by arrows;    %contained in screw and glide symmetry operators; 
    $t_1=(\frac{1}{2}\frac{1}{2}0)$, $t_2=(\frac{1}{2}\frac{1}{2}\frac{1}{2})$, and $t_3=(00\frac{1}{2})$. %\ky{change the color of arrows and the labels of corresponding symmetries, let's say, blue for C2x+t1 and mx+t1, green t2, orange t3. }
   % \blue{(TN: blue is not visible in Cy-AFM, I set red instead.)}
    %\ky{this is not good. use similar colors for t1 and t2 and very different color for t3. also make it darker to be visible.}
    %\thao{check the name of these group} \ky{change the caption as well: what you show here is spin configuration but not space group, right?  then add arrows for $m_z$ and $C_{2z}$ with (00$\frac{1}{2}$). %Exchange (a) and (b) so that the $C_y$-AFM comes first. }
    %\ky{the translation vectors in the figure are very messy. why don't you replace them by $t_1=(00\frac{1}{2})$, $t_2=..$, $t_3=..$ and put the label $t_1$ by the arrow in this figure? } 
    %\ky{can you change the color, light green to dark green, light orange to dark orange?} 
 %   \ky{Show x,y,z directions! Otherwise, the symmetry operations will lose their meaning!! Be careful, Thao!!} \ky{also, shift a bit  "(a) C-AFM)" to right to the center. reduce the length of arrows so that we can see the arrowhead. }
    }
    \label{fig:sym} 
\end{figure}

In crystalline $Pbnm$ space group, there are eight symmetry operations: \{$E$, $I$, $C_{2x}+(\frac{1}{2} \frac{1}{2} 0)$, $C_{2y}+(\frac{1}{2} \frac{1}{2} \frac{1}{2})$, $C_{2z}+(0 0 \frac{1}{2})$, $m_x +(\frac{1}{2} \frac{1}{2} 0)$, $m_y+(\frac{1}{2} \frac{1}{2} \frac{1}{2})$, $m_z+(0 0 \frac{1}{2})$\}. Introducing the time-reversal symmetry $\theta$ and adding the products of $\theta$ and these eight symmetry operations, it will be nonmagnetic $Pbnm1'$ space group having 16 symmetry operations. By considering an AFM order, the symmetry is reduced according to the spin configuration. 
When SOC is not taken into account, the spin direction does not have the physical meaning and the spin state shows just two states, {\it i.e.}, up- and down-spin states. 
To treat the non-SOC magnetic symmetry, {\it ``black-and-white group''} is sufficient~\cite{Burns_book}. 
As a matter of fact, the black-and-white group is often mixed up with the magnetic space group in the literature. Now we redefine it as follows. 
Figure~\ref{fig:sym} (a) shows the C-AFM ordering where different spin sites are shown by black and white spheres. In the black-and-white group, rotation and mirror symmetry operations do not flip the spin, but only the time-reversal symmetry operation $\theta$ flips the spins  as we turn over a black-and-white disk in the Reversi board game. Therefore, the black-and-white space group has eight symmetry operations: \{$E, I, C_{2z}+(00\frac{1}{2}),  m_z+(00\frac{1}{2}),   C_{2x}\theta+(\frac{1}{2}\frac{1}{2}0), C_{2y}\theta+(\frac{1}{2}\frac{1}{2}\frac{1}{2}),  m_x\theta+(\frac{1}{2}\frac{1}{2}0), m_y\theta+(\frac{1}{2}\frac{1}{2}\frac{1}{2})$\}; we name it  black-and-white $Pb'n'm$ space group and distinguish it from the magnetic $Pb'n'm$ space group that appeared in  Table~\ref{table:symmetry} despite the same name. 
Among those symmetry operations, 
the latter half accompanies spin-reversal symmetry $\theta$ and links the up-spin (black) sites and down-spin (white) sites,  
hence making two spin states equivalent in the electronic state.  
In the ${ k}$ space, each ${ k}$ point has the little group of the black-and-white $Pb'n'm$ group, while the translation part in screw and glide operations is dropped.  
If the considered ${ k}$ point has one of  four symmetries : \{$C_{2x}\theta$, 
$C_{2y}\theta$,
$m_x\theta$,
$m_y\theta$\} originating from the aforementioned screw and glide symmetries, it results in the spin degeneracy of the bands. Hereinafter, we call them {\it ``spin-degenerate symmetry operations''} for the AFM order.  
Other symmetry operations in the space group, \{$E$, $I$, $C_{2z}+(00\frac{1}{2}$), $m_z+(00\frac{1}{2})$\}, link the same spin sites and do not cause the spin degeneracy. Obviously, in the FM order, there are no spin-degenerate symmetry operations and hence there is no spin-degeneracy in the band structure.

\begin{table}[ht]
%\begin{minipage}[t]{1.8\columnwidth}
%\centering\def\arraystretch{1.1}
      \begin{tabular}{c|RRRRRRRR|c}
    \multicolumn{1}{C}{} & \multicolumn{8}{C}{\Gamma, {\rm X, S}} \\
{\rm C-AFM}  & E & I   & C_{2z} &  m_z  &   {C_{2x}\theta} & {C_{2y}\theta}  &  {m_x\theta} &{m_y\theta}  & ss \\
 \hline
$s$  & 1 & 1 & 1 & 1  & -1 & -1 & -1 & -1 & no\\
  \end{tabular}
  \\
      \begin{tabular}{c|RRRRRRRR|c}
               \multicolumn{1}{C}{} & \multicolumn{8}{C}{} \\
    \multicolumn{1}{C}{} & \multicolumn{8}{C}{\Gamma, {\rm{X, S}}} \\
{ $C_y$-AFM}  & E & I &  {C_{2x}}  & {m_x} &  {C_{2y}\theta}  & C_{2z}\theta &{m_y\theta} & m_z\theta & ss \\
 \hline
 $s_x$ & 1 & 1 & 1 & 1  & 1 & 1 & 1 & 1 & yes\\
 $s_y$ & 1 & 1  & -1 & -1 & -1  &  1 & -1 & 1 & no\\
 $s_z$ & 1 & 1 & -1 & -1 & 1 & -1  & 1 & -1 & no\\
  \end{tabular}
  \hspace{0pt}
%\end{minipage}
%\begin{minipage}[t]{0.9\columnwidth}
%\centering\def\arraystretch{1.1}
    \begin{tabular}{c|RRRR|c}
     \multicolumn{1}{C}{} & \multicolumn{5}{C}{} \\
%    \multicolumn{1}{C}{wrong} & \multicolumn{5}{C}{\Gamma{\rm X}} \\
%{\rm C-AFM} & E & m_z & {C_{2y}\theta}  & {m_x\theta} & ss \\
 %\hline
 % s & 1 & 1 & -1 & -1 & no\\
%
  \multicolumn{1}{C}{} & \multicolumn{4}{C}{\Gamma{\rm X}} \\
{\rm C-AFM} & E & m_z & {C_{2x}\theta}  & {m_y\theta} & ss \\
 \hline
  $s$ & 1 & 1 & -1 & -1 & no\\
  
  \multicolumn{1}{C}{} & \multicolumn{4}{C}{\text{ }} \\
%      \multicolumn{1}{C}{wrong} & \multicolumn{5}{C}{\Gamma{\rm X}} \\
%C_y{\rm -AFM}  & E & {C_{2x}} & {C_{2y}\theta} & C_{2z}\theta & ss\\
% \hline
% s_x & 1 & 1 & 1 & 1 & yes\\
% s_y & 1 & -1 & -1 & 1 & no \\
% s_z & 1 & -1 & 1 & -1 & no\\
 %
      \multicolumn{1}{C}{} & \multicolumn{4}{C}{\Gamma{\rm X}} \\
{$C_y$-AFM}  & E & {C_{2x}} & {m_{y}\theta} & m_{z}\theta & ss\\
 \hline
 $s_x$ & 1 & 1 & 1 & 1 & yes\\
 $s_y$ & 1 & -1 & -1 & 1 & no \\
 $s_z$ & 1 & -1 & 1 & -1 & no\\
  \end{tabular}
 % \hspace{0pt}
%\end{minipage}
%\begin{minipage}[t]{0.9\columnwidth}
\centering\def\arraystretch{1.1}
    \begin{tabular}{c|RR|c}
     \multicolumn{1}{C}{} & \multicolumn{2}{C}{} \\
    \multicolumn{1}{C}{} & \multicolumn{2}{C}{\Gamma{\rm S}} \\
%{\rm Non-mag} & E & m_z &  I\theta & C_{2z}\theta\\
% \hline
% s & 1 & -1 & -1 & -1 \\
%  \multicolumn{1}{C}{} & \multicolumn{4}{C}{\text{ }} \\
  {\rm C-AFM}  & E & m_z &  ss \\
 \hline
 %s\red{(thao)} & 1 & -1  & \red{no} &  \\
  $s$ & 1 & 1  & yes   \\
   \multicolumn{1}{C}{} & \multicolumn{2}{C}{} \\
%  \multicolumn{1}{C}{wrong} & \multicolumn{2}{C}{\Gamma{\rm S}} \\
%C_y{\rm -AFM} & E & C_{2z}\theta &  ss & \\
% \hline
% s_x & 1 & 1 & yes &  \\
% s_y & 1 & 1 & yes &  \\
% s_z & 1 & -1 & no &  \\
 %
   \multicolumn{1}{C}{} & \multicolumn{2}{C}{\Gamma{\rm S}} \\
{$C_y$-AFM} & E & m_{z}\theta &  ss  \\
 \hline
 $s_x$ & 1 & 1 & yes   \\
 $s_y$ & 1 & 1 & yes   \\
 $s_z$ & 1 & -1 & no   \\
  \end{tabular}
 %\hspace{0pt}
%\end{minipage}
\caption{ 
{Transformation property of the spin state $s$ and the spin components ($s_x$, $s_y$, $s_z$) at the high symmetric $\Gamma$, X, and S points and along the high symmetric $\Gamma$X and $\Gamma$S lines in the orthorhombic Brillouin zone. %in the momentum space for %nonmagnetic (grey group); 
Without SOC, C-AFM order follows black-and-white  $Pb'n'm$  space group and with SOC, $C_y$-AFM order  follows magnetic $Pbn'm'$ space group. %\blue{The spin-degenerate symmetry operations are highlighted in bold (see text). 
The possibility of spin splitting (ss) is  shown at the last column. 
%\ky{try to fit them in one column at a half page.} 
%\ky{use regular font for ss, yes, and no because they are not variables.}}
%\ky{there cannot be -1 transformation of $s$ in black-and-white group. put 1 always. }
%\ky{keep the order of symmetry operations; E, C2x, C2y, C2z, mx, my, ...}
%\ky{Add the tables for $\Gamma$ and X and S points to discuss the spin splitting at these points and the compatibility!} 
%\ky{it seems that thao still don't understand the black-and-white group; here only $\theta$ can flip the spin. if $s$ is always "1", spin splitting is allowed, right?}
}
\label{table:sym}}
\end{table}

\begin{figure}
    \centering
    \includegraphics[width=0.9\columnwidth]{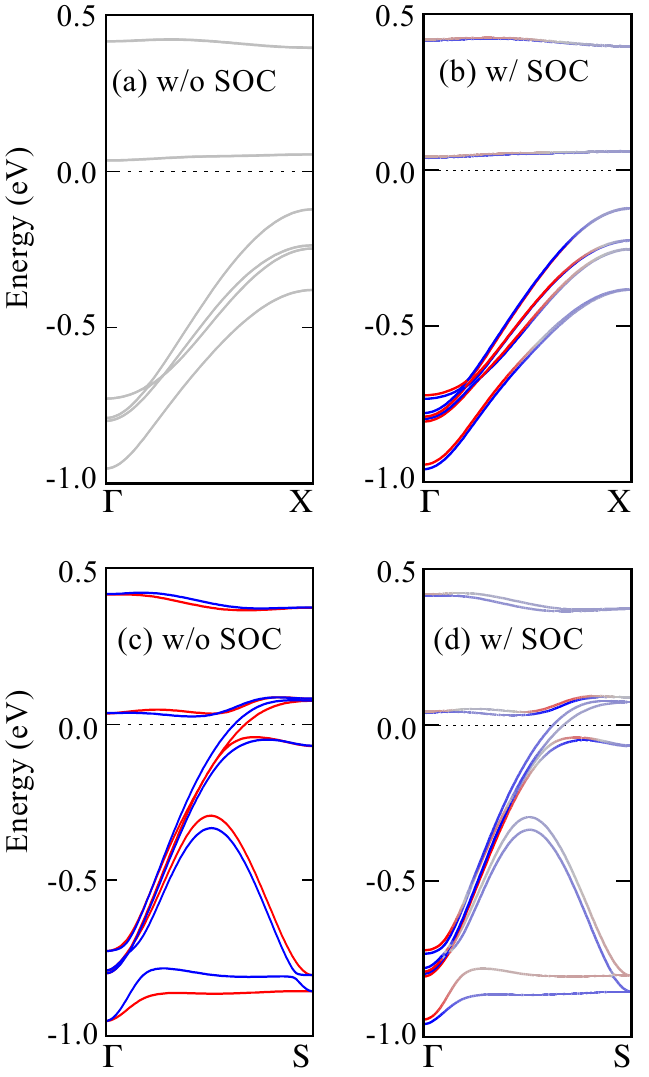}
    \caption{Calculated band structure  (a, b)  along the $\Gamma$X line and (c, d) along the $\Gamma$S line in the orthogonal Brillouin zone. In (a) and (c),  SOC is turned off and spin states are shown by red and blue colors. In (b) and (d), SOC is turned on and $s_x$ polarization is shown by red and blue colors. Fermi energy is set as zero energy. 
   % (a) and (c) Band structure along $\Gamma$X and $\Gamma$S direction of C-AFM order without SOC. 
 %  In (c) red and blue colors represent the up-spin channel and down-spin channel bands. (b) and (d) are similar to (a) and (c) but for $C_y$-AFM order with SOC. The spin polarization along the
%$x$ direction ($S_x$) is shown by red and blue colors. %\thao{change size later} 
%\ky{ replace "Sx" by "w/ SOC" in the figure.} 
%\ky{Sorry, but the SOC-splitting is very small to see. Can you limit the yrange as [-1.0:0.2] to zoom in? Can you increase k points between G and S? I guess that the anti-crossing is a fake result due to the rough k mesh. } %\thao{I notice that, I am working on more dense k-mesh. Update soon. }
%\blue{(TN: k-mesh between G and S is set to be 100, still remain anti-crossing. Maybe two bands have same IR, that make them repel each other.)}
%\ky{Now better. but you may change the font size of Energy and numbers to smaller and cut the margin. Change the blue color to gray color in (a) for all the bands because they must be spin-degenerate along GX line. } 
}
\label{fig:band_GX_GS} 
\end{figure}

Here, we demonstrate the symmetry analysis by taking band structure along the $\Gamma$X and the $\Gamma$S axes as examples; the X and S points are located at $ k$=$(\frac{1}{2}, 0, 0)$ and $(\frac{1}{2}, \frac{1}{2}, 0)$ points in the Brilloin zone, respectively. 
Table~\ref{table:sym} shows the transformation properties of spin momenta along these lines. %The symmetry operations that cause spin degeneracy are highlighted. 
Without SOC, it can be deduced that the $\Gamma$X line does not show the spin splitting while the $\Gamma$S line shows the spin splitting; the $\Gamma$S line holds $m_z$ symmetry in the little group, but it is not the spin-degenerate symmetry operation. In other words, the spin state $s$ is invariant under all the symmetry operations along the $\Gamma$S line. This was confirmed in our  band structure calculation shown in Fig.~\ref{fig:band_GX_GS} (a) and (c). It is interesting to see that the bands are spin-degenerate between the $\Gamma$ and the X points and  largely spin-split  between the $\Gamma$ and the S points without help of SOC. This splitting originates from the AFM configuration that breaks the effective $PT$ symmetry ({\it i.e.}, product of the space-inversion and time-reversal symmetries)~\cite{Yuan_Zunger2021}. 
%\blue{At the $\Gamma$, X, and S points, presence of the inversion symmetry together with the anti-unitary symmetry prevent the spin-splitting. }\ky{Make the table for these points and confirm what I am writing here!}
Remind that the spin considered here does not have any preferential direction but just black-and-white states.  

When SOC is turned on, the spin direction matters. Considering the spins as axial vectors, the mirror and rotation symmetry operations can flip the spin components ({\it e.g.}, $m_x$ mirror operation flips $s_y$ and $s_z$ but not $s_x$) as well as the time-reversal symmetry. 
By considering $C_y$-AFM ordering, the nonmagnetic $Pbnm1'$ symmetry is reduced to be $Pbn'm'$ magnetic space group with eight symmetry operations : \{$E$, $I$, $C_{2x}+(\frac{1}{2} \frac{1}{2} 0)$, $m_{x}+(\frac{1}{2} \frac{1}{2} 0)$, $C_{2y}\theta+(\frac{1}{2} \frac{1}{2} \frac{1}{2})$, $C_{2z}\theta +(0 0 \frac{1}{2})$, $m_y\theta +(\frac{1}{2} \frac{1}{2} \frac{1}{2})$, $m_{z}\theta+(0 0 \frac{1}{2}) $\}. 
%This magnetic symmetry well explains the DFT-calculated band structure taking into account SOC.
In $C_y$-AFM order, the spin momenta are collinearly aligned along the $y$ direction and slightly canted towards the $x$  direction.  As shown in Table~\ref{table:sym}, the spin components $s_x$, $s_y$, $s_z$ show different symmetry properties. Since $C_y$-AFM order has the same irreducible representation as $F_x$ order, $s_x$ is invariant under all the symmetry operations; $s_x$ 
polarization is allowed to arise everywhere in the ${ k}$ space as spin polarization in FM order. 
Unlike the AFM-induced spin splitting case, the magnitude of the  spin-splitting depends on the magnitude of SOC in this case. 
As shown in Fig.~\ref{fig:band_GX_GS} (b) and (d), the band structure exhibits the additional small band splitting by SOC on top of the non-SOC band structures (cf.  Fig.~\ref{fig:band_GX_GS} (a) and (c)). 
Along the $\Gamma$X axis, $s_x$ polarization is induced by SOC while $s_y$ and $s_z$ polarization is still zero. %\blue{(Soon after the tables are updated by thao, think of it:) The nonzero $s_x$ is allowed at the $\Gamma$ point but not allowed at the X point, the spin-splitting increases as it goes apart from the X point, similar to the Rashba effect.\cite{oguchi.J Phys Condens Matter. 2009, ky.prl2015.biiro3, ky.prb2019.bicoo3} }
%\ky{So far, I don't know why band degeneracy occurs at X point although the symmetry does not prevent $s_x$ polarization. This comes from the reason we do not consider, such as double group representation. yeah, this is band degeneracy, independent of spin degeneracy. We just do not discuss it in this paper. } 
Along the $\Gamma$S axis, $s_x$ is induced by SOC and $s_y$ is induced by AFM order, while $s_z$ is zero. %\blue{At the S point, the *** symmetry causes the degeneracy of $s_y$.}\ky{After the table is updated, think the difference between $\Gamma$ and S points, showing different band splitting!} 
Owing to the weak SOC interaction at Cr $3d$ orbital state, the SOC-induced band splitting is much smaller than the AFM-induced splitting. 

The same analysis can apply to the spin-splitting property discussed in Section D. 
Along the $\Lambda$H line, there are \{$E$, $m_x\theta$\}  symmetry operations originating from the black-and-white space group and the latter protects the spin degeneracy of the band structure without SOC; the degeneracy is lifted by turning on SOC because $s_x$ polarization is allowed by \{$E$, $m_x$\}   symmetry originating from the magnetic space group as shown in Fig. \ref{fig:spin_polarized_bands} (b) and (c). 
Along the non-symmetric $k$-line, all the spin polarization is allowed with/without SOC as shown in Fig. \ref{fig:spin_polarized_bands} (d) and (e). 
For further analysis, information on the other high-symmetric ${ k}$ points and their symmetry in the magnetic space group can be found in the Bilbao Crystallographic Server~\cite{Aroyo2011183, Aroyo_crys_mat_2006, Aroyo_Acta_Crys_2006}. 
%M. I. Aroyo, J. M. Perez-Mato, D. Orobengoa, E. Tasci, G. de la Flor, A. Kirov"Crystallography online: Bilbao Crystallographic Server" Bulg. Chem. Commun. 43(2) 183-197 (2011). Scopus Web of Science
%M. I. Aroyo, J. M. Perez-Mato, C. Capillas, E. Kroumova, S. Ivantchev, G. Madariaga, A. Kirov & H. Wondratschek "Bilbao Crystallographic Server I: Databases and crystallographic computing programs" Z. Krist. 221, 1, 15-27 (2006). doi:10.1524/zkri.2006.221.1.15
%M. I. Aroyo, A. Kirov, C. Capillas, J. M. Perez-Mato & H. Wondratschek "Bilbao Crystallographic Server II: Representations of crystallographic point groups and space groups" Acta Cryst. A62, 115-128 (2006). doi:10.1107/S0108767305040286 ]. 
We confirmed that the spin degeneracy is  symmetry-protected  along all the ${\bm k}$ paths in the band structure plot in Fig.~\ref{fig:structure} (c) without SOC.

\section{Effect of Hubbard $U$ correction}

\begin{figure}[ht]
    \centering
    \includegraphics[width=0.8\columnwidth]{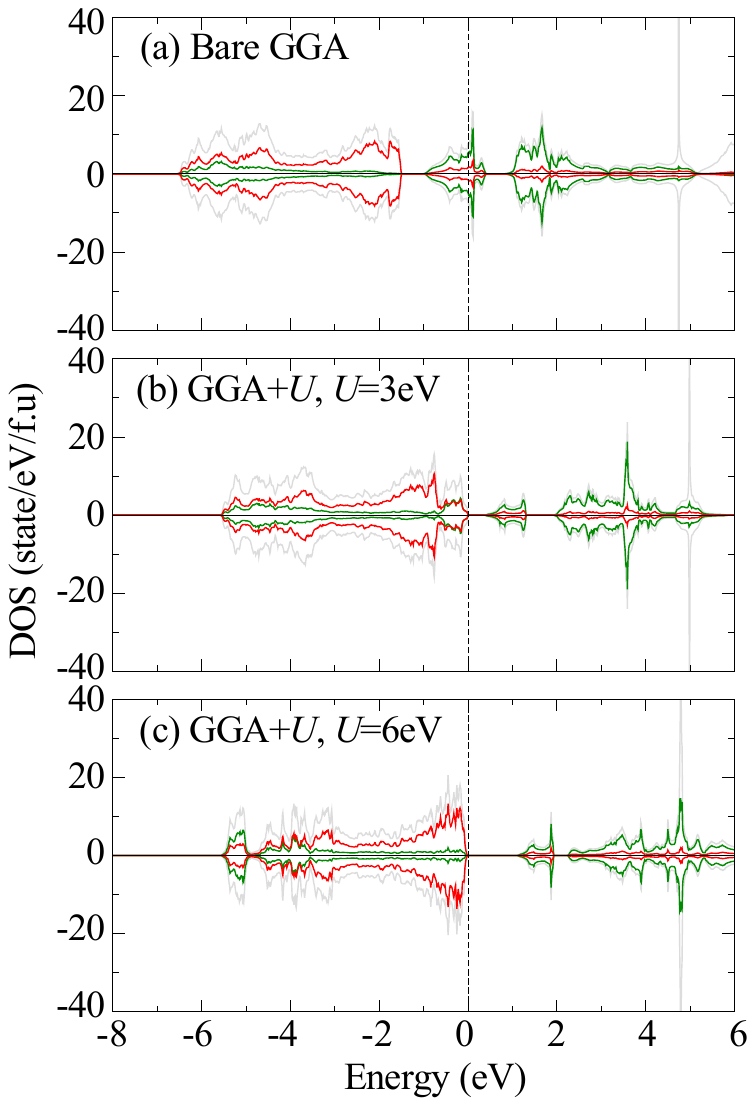}
    \caption{ Calculated total DOS (gray lines) with GGA and GGA+$U$ functionals with $U$ = 3 and 6 eV in C-AFM CaCrO$_3$ without SOC. %\ky{what is red and green?}
    Projected DOSs onto Cr-$d$ and O-$p$  orbital states are also shown by green and red lines, respectively.  
    \label{fig:DFT+U}}
\end{figure}

\begin{table}[ht]
\centering\def\arraystretch{1.1}
      \begin{tabular}{c|c c c c}
%    \multicolumn{1}{C}{} & \multicolumn{4}{C}{\Gamma, {\rm X, S}} \\
$\Delta$E (meV/f.u) & FM &  A-AFM  & C-AFM &  G-AFM  \\
 \hline
%\sout{Bare GGA(QE)}	& \sout{3.1}	& \sout{1.2} &	 0	&\sout{7.7}\\
Bare GGA	&209.2	&59.4	&{\bf 0}	&103.9\\
 GGA+$U$ ($U$=3eV)	&{\bf 0}	&12.6  &	57.5 &	17.2 \\
  GGA+$U$ ($U$=6eV)	&{\bf 0}	&61.8	&52.3	&52.9 \\
  \end{tabular}
    \caption{Relative total energy (meV/f.u) calculated with GGA and GGA+$U$ functionals with $U=$ 3 eV and $U=$ 6 eV for several spin configurations: FM, A-, C-, and G- type AFM configurations. The lowest energy is highlighted. The crystal structure (both the lattice parameters and the atomic coordinates) was relaxed under C-AFM order and then the total energy was compared by using the VASP code with 12$\times$12$\times$10 $k$-point mesh.  %\ky{Add U=6eV result, too}
    %\ky{write an apology in a reply to referees for why your values changed so much }
    }
    \label{tab:energy_gga+u}
\end{table}

%\ky{Thao, Shortly summarize GGA+U result here. Add the magnetic stability results.} 
%\sout{For completion} \blue{(?)}, 
In strongly correlated materials, 
the Hubbard $U$ correction of DFT+$U$ approach is widely used to reinforce the on-site Coulomb repulsion for $3d$ or $4f$ localized orbital states~\cite{DudarevPRB1998}. %\cite{DedarevPRB1998}} 
%Here we provide the dependence of electronic properties on Hubbard $U$ correction in \cacro. 
Figure~\ref{fig:DFT+U} shows the change in the electronic state in CaCrO$_3$ with GGA and GGA+$U$ potentials with various $U$ values.  
At bare GGA level (with $U$ = 0 eV), the Fermi level crosses the Cr-$t_{2g}$ orbital state showing the good metallic state. 
 As increasing the $U$ value, we found the band-gap opening with $U \gtrsim $ 2.5 eV.  %\ky{write the exact value. 
%, the insulating phase is stabilized with a band gap of about 0.35eV. 
The band gap increases from 0.3 eV at $U$ = 3 eV to 1.0 eV at $U$ = 6 eV. This is consistent with the result reported by Streltsov {\it et al.}; the electronic state is metallic at LSDA level and insulating at LSDA+$U$ level~\cite{Streltsov_cacro3_dft}. Nevertheless, %DFT+U formalism tends to drive this system away from the metallic regime, which 
the GGA+$U$ result is inconsistent with the experimental observations of metallicity in CaCrO$_3$~\cite{Zhou2006_cacro3}. %For the sake of consistency with experiment, 
%GGA functional is better for the description of the electronic properties as well as the electric conductivity.
%
Table~\ref{tab:energy_gga+u} shows the relative total energy for four magnetic configurations with several $U$ values. The bare GGA result  correctly reproduces the C-AFM ground state, consistent with the experiment, while  GGA+$U$ results show the wrong solution as the FM ground state. In fact, the Hubbard $U$ correction  often overestimates the tendency toward ferromagnetism as already noted by Terakura {\it et al.} and by Picozzi {\it et al.} for orthorhombic rare-earth manganites~\cite{TerakuraPRB1996, PicozziPRB2006}. 
Considering the better description of the metallic state and the magnetic stability, the bare GGA functional was used in our DFT calculation. % for CaCrO$_3$. 
% To investigate the role of Hubbard $U$ correction on the magnetic stability, we perform the total energy different calculations using GGA and GGA+$U$ functional. In 

%Overall, due to the aforementioned reasons, we decided to use bare GGA functional for this study for the consistency between DFT calculation and experimental data.

%\section{References}
\bibliography{mybibitem.bib}
\bibliographystyle{apsrev4-2} 
%\ky{Give the title of Vanderbilt's book}
\end{document}